\documentclass[onecolumn,draftclsnofoot,12pt]{IEEEtran}

\usepackage{enumerate}

\usepackage{amsmath,amsthm}
\usepackage[noend]{algpseudocode}
\usepackage{float}
\usepackage{hyperref}
\usepackage{color}
\usepackage{makeidx}
\usepackage{bbm}
\usepackage{graphicx}
\usepackage{lipsum}
\usepackage{soul}
\usepackage{tabularx}
\usepackage{dsfont}
\usepackage{amsmath}
\usepackage{algorithm}
\usepackage[noend]{algpseudocode}

\usepackage{amsfonts}
\usepackage{times}
\usepackage{graphicx}
\usepackage{latexsym}
\usepackage{dsfont}
\usepackage{amssymb}
\usepackage{amsmath}
\usepackage{cite}
\usepackage{verbatim}
\usepackage{subfigure}

\newcommand{\figref}[1]{{Fig.}~\ref{#1}}

\def\bb0{{\mathbb{0}}}

\def\ba{{\mathbf{a}}}
\def\bb{{\mathbf{b}}}

\def\bm{{\mathbf{m}}}

\def\bw{{\mathbf{w}}}

\def\b0{{\mathbf{0}}}

\def\bA{{\mathbf{A}}}
\def\bB{{\mathbf{B}}}

\def\bR{{\mathbf{R}}}

\def\bbE{{\mathbb{E}}}

\def\cA{\mathcal{A}}

\def\cN{\mathcal{N}}

\def\sf0{{\mathsf{0}}}

\usepackage{epstopdf}

\newcommand{\sref}[1]{{Section}~\ref{#1}}

\newcommand{\pinv}[1]{\ensuremath{#1^{\dagger}}}

\DeclareMathOperator*{\argmax}{arg\,max}

\def\rm{\mathrm}

\newcommand{\subto}{\operatorname{s.t.}}

\begin{document}

\title{Neural Networks Based  Beam Codebooks: Learning mmWave Massive MIMO Beams that Adapt to  Deployment  and Hardware}

\author{Muhammad Alrabeiah, Yu Zhang and Ahmed Alkhateeb \thanks{Muhammad Alrabeiah, Yu Zhang and Ahmed Alkhateeb are with Arizona State University (Email: malrabei, y.zhang, alkhateeb@asu.edu). This work is supported by the National Science Foundation under Grant No. 1923676. A conference version of this paper has been published in \cite{Zhang2020}}}
\maketitle

\begin{abstract}

Millimeter wave (mmWave) and massive MIMO systems are intrinsic components of 5G and beyond. These systems rely on using beamforming codebooks for both initial access and data transmission. Current beam codebooks, however, generally consist of a large number of narrow beams that scan all possible directions, even if these directions are never used. This leads to very large training overhead. Further, these codebooks do not normally account for the hardware impairments or the possible non-uniform array geometries, and their calibration is an expensive process. To overcome these limitations,  this paper develops an efficient online machine learning framework that learns how to adapt the codebook beam patterns to the specific deployment, surrounding environment, user distribution, and hardware characteristics. This is done by designing a novel \textit{complex-valued neural network} architecture in which the neuron weights directly model the  beamforming weights of the analog phase shifters, accounting for the key hardware constraints such as the constant-modulus and  quantized-angles. This model learns the codebook beams through  online and self-supervised training avoiding the need for explicit channel state information. This respects the practical situations where the channel is either unavailable, imperfect, or hard to obtain, especially in the presence of hardware impairments. Simulation results highlight the capability of the proposed solution in learning environment and hardware aware beam codebooks, which can significantly reduce the training overhead, enhance the achievable data rates, and improve the robustness against possible hardware impairments.
\end{abstract}


\section{Introduction} \label{intro}


Millimeter wave (mmWave) MIMO is an essential ingredient of the future wireless communication networks---from 5G to IEEE 802.11ay and beyond \cite{Ghasempour2017,Andrews2014,Roh2014,Giordani2019a}. These systems use large antenna arrays to obtain enough beamforming gains and guarantee sufficient receive signal power. Due to the cost and power consumption of the mixed-signal circuits at the high frequency bands, though, fully-digital transceiver architectures that assign an RF chain per antenna are not feasible \cite{Alkhateeb2014d}.  Instead, these mmWave systems resort to analog-only or hybrid analog/digital architectures \cite{Alkhateeb2014} to implement the beamforming/combining functions. Further, because of the hardware constraints on these large-scale MIMO systems and the difficulty in channel estimation and feedback, they typically adopt pre-defined single-lobe beamforming codebooks (such as DFT codebooks \cite{Hur2013}) that scan all possible directions for both initial access and data transmission. Examples of using these codebooks include the Synchronization Signal Block (SSB) beam sets and Channel State Information Reference Signal (CSI-RS) codebooks in 5G \cite{Giordani2019}, and hierarchical beam patterns in IEEE 802.11ad  \cite{11ad}. The classical beam-steering codebooks, however, have several drawbacks: (i) They incur high beam training overhead by scanning all possible directions even though many of these directions may never be used, (ii) they normally have single-lobe beams which may not be optimal, especially in non-line-of-sight (NLOS) scenarios, and (iii) they are typically predefined and do not account for possible hardware imperfections (such as phase mismatch or arbitrary array geometries) with very expensive calibration processes \cite{Moon2019}. \textbf{To overcome these limitations, we propose a novel online machine learning framework that learns how to adapt the codebook beam patterns to the surrounding environment, the user distribution, and the given hardware of the specific base station deployment}---building what we call environment and hardware aware beam codebooks. 

\subsection{Prior Work:}

Designing MIMO beamforming codebooks has been an important research and development topic for a long time at both academia and industry \cite{Jindal2006,Au-yeung2007,Huang2009,Raghavan2007a,Love2005b,Raghavan2008a,Lee2009,IEEE11n2012}. The motivation for all this prior work has been mainly to enable efficient limited-feedback operation in MIMO systems. For example, the authors of \cite{Jindal2006,Au-yeung2007} investigated the design of beamforming codebooks for MISO communication systems with Rayleigh channels. The same problem was then considered in \cite{Huang2009,Raghavan2007a} for spatially and temporally correlated channels. For systems with multiple antennas at both the transmitters and receivers,  \cite{Love2005b,Raghavan2008a} developed precoding/combining codebooks and analyzed the system performance under various channel models. The use of beam codebooks have been also adopted by several cellular and wireless LAN standards \cite{Lee2009,IEEE11n2012}. The codebook approaches in \cite{Jindal2006,Au-yeung2007,Huang2009,Raghavan2007a,Love2005b,Raghavan2008a,Lee2009,IEEE11n2012}, however, were generally designed to optimize the feedback of small-scale MIMO   and are hard to  extend to massive MIMO systems without the requirement of huge codebook sizes and large training overhead. Further, the codebooks in  \cite{Jindal2006,Au-yeung2007,Huang2009,Raghavan2007a,Love2005b,Raghavan2008a,Lee2009,IEEE11n2012} adopted fully-digital architectures and did not consider the hardware constraints at the transmitter/receiver arrays  which could highly affect the design of these codebooks. Incorporating these constraints is essential for the development of efficient mmWave MIMO codebooks.

For mmWave systems, \cite{Wang2009,Hur2013,Alkhateeb2014,Alkhateeb2016d}  developed a set of new beamforming codebooks for analog-only and hybrid analog/digital architectures. In \cite{Wang2009}, narrow-beam codebooks were developed to aid the beam training processing mmWave systems. The narrow beams, however, may lead to large training overhead. This motivated designing hierarchical codebooks in \cite{Hur2013,Alkhateeb2014}  that consist of different levels of beam widths. For  frequency selective channels, \cite{Alkhateeb2016d} developed iterative hybrid analog/digital beamforming codebooks. The codebooks in \cite{Wang2009,Hur2013,Alkhateeb2014,Alkhateeb2016d}, however, have several limitations. First, they were generally designed for unconstrained architectures and then approximated for these constraints, i.e., they were not particularly optimized for these hardware constraints. Second, they were mainly designed to have single-lobe narrow beams that cover all the angular directions and are not adaptive to the particular deployment characteristics (surrounding environment, user distributions, etc.), which requires large training overhead. Further, these codebooks assumed fully-calibrated uniform arrays and experience high distortion in practical hardware with fabrication impairments. All that motivated the development of environment and hardware aware codebook learning strategies, which is the focus of this paper.

\subsection{Contribution:}

In this paper, we consider hardware-constrained large-scale MIMO systems and propose an artificial neural network based framework for learning environment and hardware adaptable  beamforming codebooks. The main contributions of the paper can be summarized as follows
\begin{itemize}
	\item First, we design a supervised machine learning model that can learn how to adapt the patterns of the codebook beams based on the surrounding environment and user distribution. This is done by developing a novel \textit{complex-valued neural network} architecture in which the weights directly model the beamforming/combining weights of the analog phase shifters. The proposed model accounts for the key hardware constraints such as  the phase-only, constant-modulus, and quantized-angle constraints \cite{Alkhateeb2014d}. The training process was designed to approach the performance of  equal-gain transmission/combining \cite{Love2003}, which is the upper bound of the analog-only beamforming solutions. 
	
	\item Then, we develop a second neural network architecture that relies on online and self-supervised training and avoids the need of explicit channel state information. This respects the practical situations where the channel state information is either unavailable, imperfect, or hard to obtain, especially in the presence of hardware impairments. The developed architecture learns in an online and self-supervised fashion how to adapt the codebook beam patterns to suit the surrounding environment, user distribution,  hardware impairments, and unknown antenna array geometry. 		
	
	\item Finally, we extensively evaluate the performance of the proposed codebook learning approaches based on the publicly-available DeepMIMO dataset \cite{DeepMIMO}. These experiments adopt both outdoor and indoor wireless communication scenarios at different signal-to-noise ratios (SNRs) and codebook sizes. Further, this evaluation is done  both for uniform-perfect arrays and for arrays with arbitrary geometries and hardware impairments. These experiments provide a comprehensive evaluation of the proposed codebook learning approaches.  
\end{itemize}

The simulation results verify the effectiveness of the proposed solutions in providing the sought-after environment and hardware awareness. In particular, the proposed solutions show significant improvements compared to classical beam-steering codebooks in several cases: (i) For arbitrary user distributions in which our approaches learn how to adapt the beams to focus on where the users are and significantly reduce the required beam training overhead,  (ii) for NLOS scenarios with multiple equally-strong paths where the developed codebook learning solutions learn multi-lobe beams that achieve much higher data rates, and (iii) for arrays with hardware impairments or unknown  geometries, where our neural networks learn how to adapt the beam patterns for the given arrays and  mitigate the impact of hardware impairments. All that highlights a promising direction where machine learning can be integrated into the communication systems to develop deployment and hardware specific beam codebooks.

\textbf{Notation}: $\bA$ is a matrix, $\ba$ is a vector, $a$ is a scalar, and $\cA$ is a set. $\bA^T$, $\bA^H$,  $\pinv{\bA}$ are its transpose, Hermitian,  and pseudo-inverse respectively. $[\bA]_{m,n}$ is the element in the $m$th row and $n$th column. $\mathrm{diag}(\ba)$ is a diagonal matrix with the entries of $\ba$ on its diagonal.  $\bA \otimes \bB$ is the Kronecker product of $\bA$ and $\bB$, and $\bA \circ \bB$ is their Khatri-Rao product. $\cN(\bm,\bR)$ is a complex Gaussian random vector with mean $\bm$ and covariance $\bR$. $\bbE\left[\cdot\right]$ is used to denote expectation.

\clearpage
\section{System and Channel Models} \label{sec:System}
\begin{figure}[t]
	\centering
	\includegraphics[width=.8\linewidth, height=230pt]{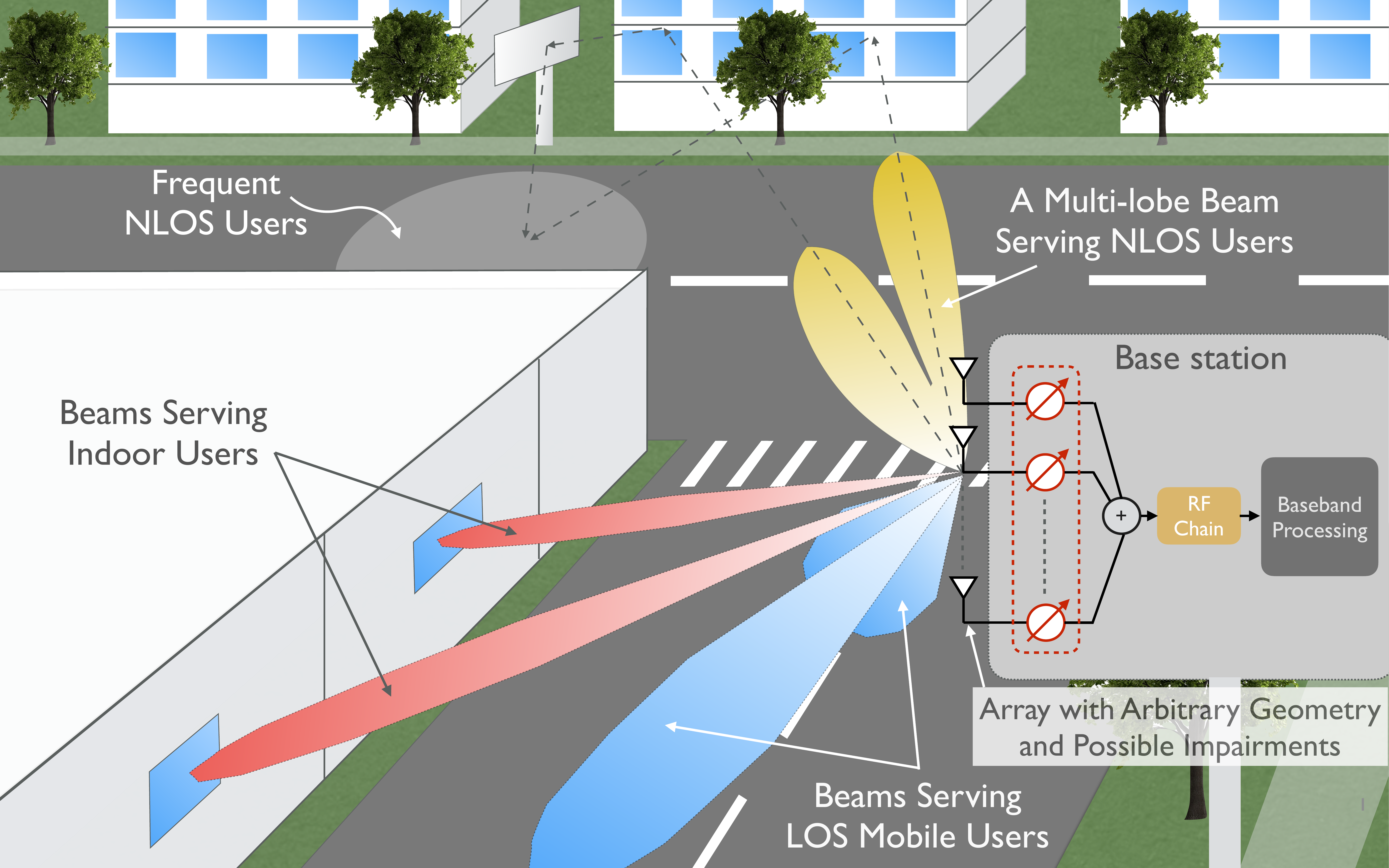}
	\caption{The adopted system model where a base station of $M$ antennas can communicate with LOS or NLOS users using a beam codebook. The proposed machine learning model in this paper learns how to efficiently adapt the codebook  beams based on the given deployment, user distributions, and hardware characteristics.}
	\label{sys_model}
\end{figure}

In this section, we describe in detail our adopted system and channel model. Further, we describe how the model considers arbitrary arrays with possible hardware impairments. 

\subsection{System Model}

We consider the communication system shown in \figref{sys_model} where a base station (BS) with $M$ antennas is deployed in a certain environment and is capable of serving both the LOS and NLOS mobile users in this environment. For simplicity, we assume that the users have single antennas. The proposed solutions in this paper, however, could be extended to the case with multi-antenna users.  Next, considering the uplink transmission, if the user transmits a symbol $s \in\mathbb{C}$,  then the received signal at the BS after combining can be expressed as
\begin{equation}\label{sys}
y = {\mathbf w}^H{\mathbf h} s + {\mathbf w}^H{\mathbf n},
\end{equation}
where the transmitted symbol satisfies the  average power constraint $\mathbb{E}\left[|s|^2\right]=P_s$ and ${\mathbf n}\sim\mathcal{N}_\mathbb{C}\left(0, \sigma_n^2{\bf I}\right)$ is the receive noise vector at the BS. The $M \times 1 $ vector  ${\mathbf h}\in\mathbb{C}^{M\times 1}$ denotes the uplink channel between the mobile user and the BS antennas and ${\bf w}$ represents the BS combining vector. Given the cost and power consumption of the mixed-signal components at the mmWave frequencies, it is hard to dedicate an RF chain for each antennas and apply fully-digital precoding/combining at mmWave massive MIMO systems \cite{Alkhateeb2014d}. Alternatively, mmWave base stations adopt analog-only or hybrid analog-digital beamforming approaches that move all or some of the beamforming/combining processing to the RF domain  \cite{Alkhateeb2014d,Roh2014}. To account for that, we assume that the BS employs an analog-only architecture where the beamforming/combining is implemented using a network of phase shifters as depicted in \figref{sys_model}. With this architecture, the combining vector $\bw$ can be written as
\begin{equation}\label{Analog}
  {\bf w} = \frac{1}{\sqrt{M}}\left[ e^{j\theta_1}, e^{j\theta_2}, \dots, e^{j\theta_M} \right]^T,
\end{equation}
which can only perform phase shift to the signal received by each antenna.

\subsection{Channel Model} \label{subsec:channel}

We adopt a general geometric channel model for ${\mathbf h}$ \cite{Alrabeiah2019,Li2019}. Assume that the signal propagation between the mobile user and the BS consists of $L$ paths. Each path $\ell$ has a complex gain $\alpha_\ell$ (that includes the path-loss) and an angle of arrival $\phi_\ell$. Then, the channel can be written as
\begin{equation}\label{channel}
  {\mathbf h}=\sum\limits_{\ell=1}^{L}\alpha_\ell{\mathbf a}(\phi_\ell),
\end{equation}
where ${\bf a}(\phi_\ell)$ is the array response vector of the BS. The definition of ${\bf a}(\phi_\ell)$ depends on the array geometry and hardware impairments which we discuss in the following subsection.

\subsection{Arbitrary Array Geometry and Hardware Impairments} \label{sec:impairments}

Most of the prior work on mmWave signal processing has  assumed uniform antenna arrays with perfect calibration and ideal hardware \cite{Hur2013,Wang2009,Alkhateeb2014,Alkhateeb2014d}. In this paper, we consider a more general antenna array model that accounts for arbitrary geometry and hardware imperfections. We show that our online beam codebook learning approaches can efficiently learn beam patterns for these arrays and adapt to their particular characteristics. This leads to several advantages for these systems since (i) there are  scenarios where designing arbitrary arrays is needed, for example, to  improve the angular resolution  or enhance the direction-of-arrival estimation performance \cite{Pal2010,Rubsamen2009},  (ii) the fabrication process of large mmWave arrays normally has some imperfections, and (iii) because the calibration process of the  mmWave phased arrays is an expensive process that requires special high-performance RF circuits \cite{Moon2019}. 
While the codebook learning solutions that we develop in this paper are general for various kinds of arrays and hardware impairments, we evaluate them in \sref{sec:results} with respect to two main characteristics of interest, namely  non-uniform spacing and phase mismatch between the antenna elements.  For linear arrays, the array response vector can be modeled to capture these characteristics as follows
\begin{equation}\label{ARV-cor}
	{{\bf a}}(\phi_\ell) = \left[ e^{j\left(kd_1\cos(\phi_\ell) + \Delta\theta_1\right)}, e^{j\left(kd_2\cos(\phi_\ell) +\Delta\theta_2\right)},\dots, e^{j\left(kd_M\cos(\phi_\ell) + \Delta\theta_M\right)} \right]^T,
\end{equation}
where $d_m$ is the position of the $m$-th antenna, and $\Delta\theta_m$ is the additional phase shift incurred at the $m$-th antenna (to model the phase mismatch). Without loss of generality, we assume that $d_m$ and $\Delta\theta_m$ are fixed yet {unknown} random realizations, obtained from the distributions $\mathcal{N}\left((m-1)d, \sigma_d^2\right)$ and $\mathcal{N}\left(0, \sigma_p^2\right)$, respectively.


\section{Problem Definition} \label{sec:Prob}
In this paper, we investigate the design of mmWave beamforming codebooks that are adaptive to the specific deployment (surrounding environment, user distribution/traffic, etc.) and the given hardware (array geometry, hardware imperfections, etc.), as shown in \figref{sys_model}.  Next, we formulate the beam codebook optimization problem before showing in Sections \ref{sup_sol}-\ref{self-sup_sol} how neural network based machine learning can provide efficient approaches for learning adaptive codebooks.  Given the system and channel models described in \sref{sec:System}, the SNR after combining for user $u$ can be written as
\begin{equation}\label{single_snr}
  \mathsf{SNR}_u = \frac{\left|{\bf w}^H{\bf h}\right|^2}{\left|{\bf w}\right|^2} \rho, 
\end{equation}
with $\rho=\frac{P_s}{\sigma_n^2}$. If the combining vector $\bw$ is selected from a codebook $\mathcal{W}$, with cardinality $\left|\mathcal W \right|=N$, then, the maximum achievable SNR for use $u$ is obtained by the exhaustive search over the beam codebook as 
\begin{equation}
\mathsf{SNR}^\star_u= \rho \max_{\bw \in \mathcal W}  {\left|{\bf w}^H{\bf h}\right|^2},
\end{equation}
where we set $\left|\bw\right|^2=1$ as these combining weights are implemented using only phase shifters with constant magnitudes of $1/\sqrt{M}$, as described in \eqref{Analog}. Our objective in this paper is to design the codebook $\mathcal W$ to maximize the  SNR  averaged over the candidate set of user channels $\boldsymbol{\mathcal{H}}$, which are the channels of the candidate users in the environment surrounding the deployed BS. This problem can then be written as 
\begin{align}\label{Prob-0}
\mathcal{W}_{\mathsf{opt}} = & \argmax\limits_{\mathcal{W}}  ~ \sum_{{\bf h}\in\boldsymbol{\mathcal{H}}} \left( \max\limits_{\substack{\bw_n \in \mathcal{W}, \\ n=1, ..., N }}\left| {\bf w}_n^H{\bf h} \right|^2 \right), \\
& \subto ~ ~ \hspace{20pt} \left| \left[\bw_n\right]_m\right| = \frac{1}{\sqrt{M}}, ~ \forall n=1,...,N, m=1, ..., M, \label{unit-0}
\end{align}
where the constraint in \eqref{unit-0} is imposed to uphold the phase-shifters constraint, i.e., the analog beamformer can only perform phase shifts to the received signal but is not capable of adapting the gain. It is worth mentioning here that while we are focusing on receive beamforming design in this paper, the same solution can be used for transmit codebook design by acquiring SNR feedback from the users.

The objective of problem \eqref{Prob-0} is to find the beam codebook that maximizes the average SNR gain for all the candidate users. Since we only have a finite beamforming codebook, which is far less than the number of users, it is impossible to achieve the maximum combining SNR for each user (which is given by the equal-gain combining \cite{Love2003}). In this sense, we might expect to find a codebook such that each beamformer serves a group of users that share  similar channels. 
Due to the large number of channels in $\boldsymbol{\mathcal{H}}$ as well as the non-convex constraints \eqref{unit-0}, problem \eqref{Prob-0} in general is very hard to solve by using the classical optimization methods and beamforming design approaches \cite{Boyd2004,Love2008,Alkhateeb2014,Alkhateeb2016d}. Therefore, and motivated by the powerful learning and optimization capabilities of neural networks, we consider leveraging neural network based machine learning  to efficiently solve the optimization problem \eqref{Prob-0}. Depending on whether the channel state information is available or not, two different machine learning frameworks are designed, namely  supervised and self-supervised solutions, in Sections \ref{sup_sol} and \ref{self-sup_sol} to learn  beam codebooks that adapt to the given deployment and hardware---generating what we call environment and hardware aware beam codebooks. 

\section{Supervised Machine-Learning Solution}\label{sup_sol}

Designing environment-aware mmWave beam codebooks requires an adaptive and data-driven process. Data collected from the environment surrounding a base station, like channels and/or user-received power, is a powerful source of information as it encodes information on user distributions and users' multi-path signatures. Such data could be used to tailor the beamforming codebook to those users and that environment. The challenge here is the need for a system capable of sifting through the data, analyzing it, and designing the codebook in a manner that respects the phase-shifter constraint. This clearly calls for a system with a sense of intelligence.

This section addresses that challenge and proposes an elegant solution that is environmentally adaptable, data-driven, and hardware compatible. In its core, this solution relies on machine learning and, in particular, artificial neural networks \cite{DeepLearning}. It follows a supervised learning approach to analyze the channel structure and learn the phases of the \textit{suitable} beamforming vectors. Its elegance stems from the way it learns the codebook; the weights of the neural network directly relate to the angles of the phased arrays, making them the actual parameters of the network. Therefore, during every training cycle (forward and backward passes) the codebook will be updated directly.

\begin{figure}[t]
	\centering
	\includegraphics[width=.7\linewidth]{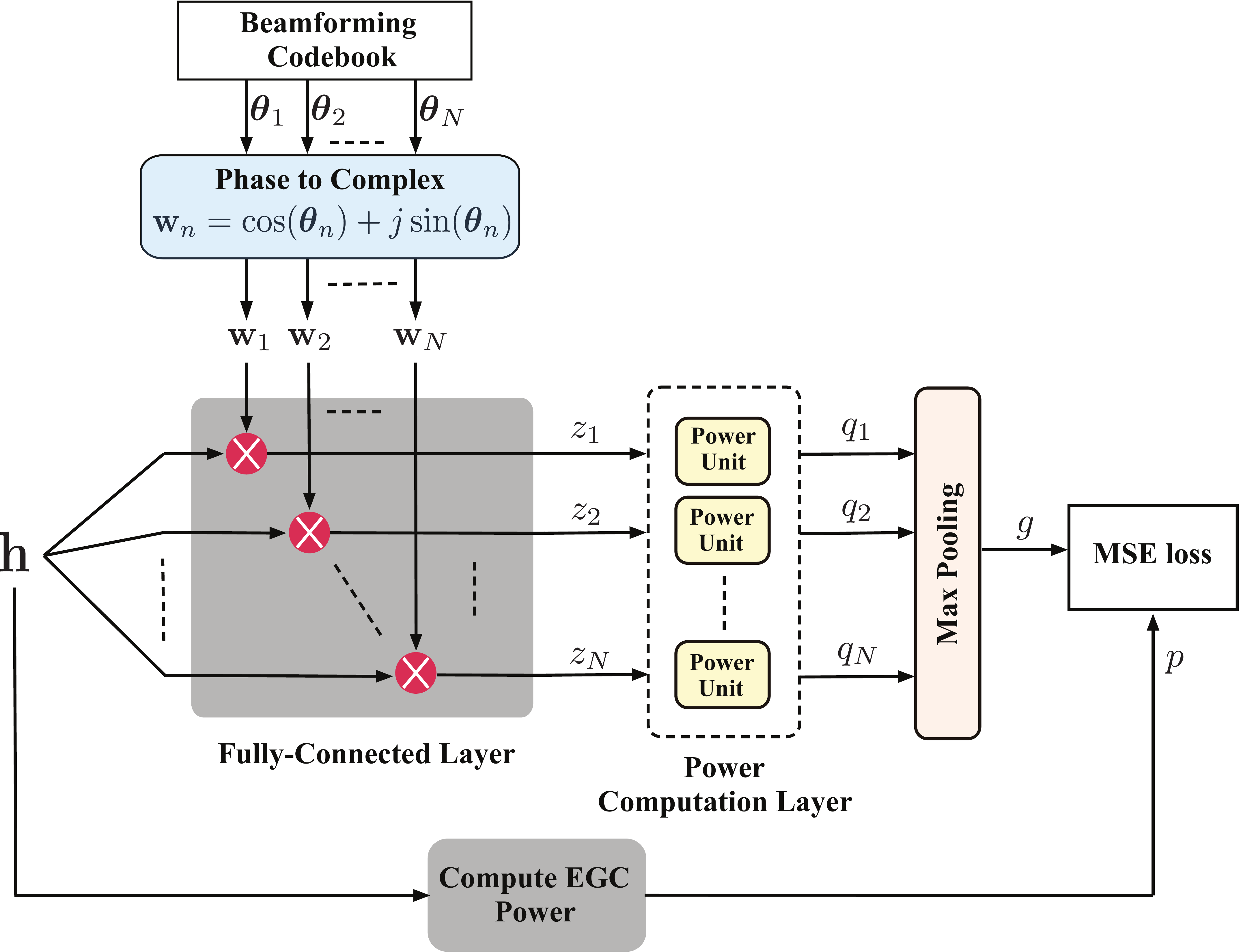}
	\caption{This schematic shows the overall architecture of the neural network used to learn the beamforming codebooks. It highlights the network architecture and the auxiliary components, equal-gain-combining and MSE-loss units, used during the training process. It also gives a slightly deeper dive into the inner-workings of the cornerstone of this architecture, the complex-valued fully-connected layer.}
	\label{Arch}
\end{figure}

\subsection{Model Architecture}\label{sup_arch}

Before going into the details of how a codebook is learned, it is important to explain the architecture of the proposed neural network and its relation to the optimization problem in \eqref{Prob-0}. This architecture consists of three main components, as depicted in \figref{Arch}. Those components are the complex-valued fully-connected layer, the power-computation layer, and finally the max-pooling layer. A forward pass through these three layer is equivalent to evaluating the cost function of \eqref{Prob-0} over a single channel $\mathbf h$.

\subsubsection{Complex-valued fully-connected layer}\label{subsub:comp_val_fc}

The first layer consists of $N$ neurons that are capable of performing complex-valued multiplications and summations. Each neuron, as shown in \figref{Arch}, learns one beamforming vector and performs inner product with the input channel vector. Formally, this is described by the following matrix multiplication
\begin{align}\label{fc_1}
  \mathbf z &= \mathbf W ^H \mathbf h,
\end{align}
where $\mathbf W = [ \mathbf w_1,\dots,\mathbf w_N]\in \mathbb C^{M\times N} $ is the beamforming codebook, $(.)^H$ is the conjugate transpose (Hermitian) operation, $\mathbf h$ is a user's channel vector, and $\mathbf z \in \mathbb C^{N\times 1}$ is the vector of the combined received signal. This equation could be re-written in the following block matrix form
\begin{align}\label{fc_2}
	\left[ \begin{array}{l}
		\mathbf z^r \\
		\mathbf z^{im}
	\end{array}\right]
	&= \left[\begin{array}{cc}
  	\mathbf W^r & \mathbf {-W}^{im} \\
  	\mathbf W^{im} & \mathbf W^{r}
  \end{array} \right]^T
  \left[ \begin{array}{l}
  	\mathbf h^r \\
  	\mathbf h^{im}
  \end{array}\right],
\end{align}
where $\mathbf z^r,\mathbf z^{im}\in \mathbb R^N$ are the real and imaginary parts of $\mathbf z$, $\mathbf W^r,\mathbf W^{im} \in \mathbb R^{M\times N}$ are matrices containing the real and imaginary components of the elements of $\mathbf W$, and, finally, $\mathbf h^r,\mathbf h^{im} \in \mathbb R^M$ are the real and imaginary components of the channel vector $\mathbf h$. What is interesting about \eqref{fc_2} is that it provides a peek behind the curtains to the inner-workings of the complex-valued fully-connected layer.

Contrary to the norm in designing neural networks, the elements of the beamforming matrix $\mathbf W$ are not the weights of the fully-connected layer. Instead, they are derived from the actual neural network weights, which are the phased arrays making up the beamforming codebook. This is done through an embedded layer of phase-to-complex operations, as shown in \figref{Arch}. This layer transforms the phased arrays into unit-magnitude complex vectors by applying elements-wise $cos$ and $sin$ operations and scale them by $1/\sqrt{M}$ as follows
\begin{align}\label{eq:ph_to_weight}
  \mathbf W &= \frac{1}{\sqrt{M}}\left( \cos\left( \mathbf \Theta\right) + j*
  \sin\left( \mathbf \Theta \right) \right),
\end{align}
where $\mathbf \Theta = \left[ \boldsymbol{\theta}_1, \dots, \boldsymbol{\theta}_{N} \right]$ is an $M \times N$ matrix of phased arrays, and $\boldsymbol{\theta_n} = [\theta_{1n}, \dots, \theta_{Mn}]^T$, $\forall n \in \{1,\dots, N\}$ is a single phase vector. The use of this embedded layer is the network's way of learning beamforming vectors that \textbf{respect the phase shifter constraint}.

\subsubsection{Power-computation layer}

The output of the complex-valued fully-connected layer feeds into the power-computation layer. It performs element-wise absolute square operation and outputs a real-valued vector ${\bf q}$ given by
\begin{equation}\label{powOut}
  {\bf q} = \left[q_1, q_2, \dots, q_N\right]^T = \left[ |z_1|^2, |z_2|^2, \dots, |z_N|^2 \right]^T,
\end{equation}
which has the received power of each beamformer in the codebook.

\subsubsection{Max-pooling layer}

The power of the best beamformer is, finally, found by the last layer, the max-pooling layer. It performs the following element-wise $\max$ operation over the elements of $\mathbf q$
\begin{equation}\label{maxOut}
  g = \max\left\{q_1, q_2, \dots, q_N\right\},
\end{equation}
and outputs $g$, which is the power of the best beamformer. This value is used to assess the quality of the codebook by comparing it to a desired receive power value. The details on what this desired value is and how the quality is assessed are detailed in the following subsection.

\subsection{Learning Codebooks}\label{sec:sup_learning}

With the neural network architecture in mind, it is time to delve into the details of how a codebook is learned. This first proposed solution, as its name states, follows a supervised learning approach. In such approach, a machine learning model is trained using pairs of inputs and their desired responses, which constitute the training dataset.

\subsubsection{Desired response}\label{sec:desir_reso}

For the beamforming problem in hand, the inputs to the model are the users' channels as they are the communication quantity that drives the beamforming design process. As training targets, there are many possible desired responses that could be used, and the choice between them should be made based on what the models needs to learn. In this paper, equal gain combining is adopted as the desired response. This choice is based on the fact that equal gain combining respects the phase shifters constraint. It is the beamforming that achieves optimal SNR performance when there are no restrictions on the codebook size. Further, equal gain combining constitutes an upper bound for the received power of fully-analog transceivers. The equal-gain combining beamformer is obtained using the phase component of every user's channel as follows
\begin{equation}\label{EGClabel}
  {\bf w}_\mathrm{EGC} = \frac{1}{\sqrt{M}}\left[ e^{\angle h_1}, e^{\angle h_2}, \dots, e^{\angle h_M} \right]^T,
\end{equation}
where $\angle$ stands for the phase of a complex number. Using equal gain combining beamformers, the desired response for each user can be computed as follows
\begin{equation}\label{EGCgain}
  p = \left| {\bf w}_\mathrm{EGC}^H{\bf h} \right|^2 = \frac{1}{M}\left\|{\bf h}\right\|_1^2,
\end{equation}
where $\|\cdot\|_1$ is the $L1$ norm. Putting the users' channels and their equal-gain combining gains together provides the training dataset $\mathcal S_{t}$. 

\subsubsection{Model background training}\label{sec:back_trn}

Using the set $\mathcal S_t$, the model is trained in the background by undergoing multiple forward-backward cycles. In each cycle, a \textit{mini-batch} of complex channel vectors and their equal-gain combining responses is sampled from the training set. The channels are fed sequentially to the model and a forward pass is performed as describe in Section \ref{sup_arch}. For each channel vector in the batch, the model combines it with the currently available $N$ beamforming vectors and outputs the power of the best beamformer for that channel. The quality of the best combiner is assessed by measuring how close its beamforming gain to that of the channel equal-gain combiner, obtained by \eqref{EGCgain}. A Mean-Squared Error (MSE) loss is used as a metric to assess the quality of the codebook over the current mini-batch. Formally, it is defined as
\begin{equation}\label{MSE-loss}
  \mathcal L = \frac{1}{B} \sum_{b=1}^B (g_b - p_b)^2,
\end{equation}
where $g_b$ is the output of the max-pooling layer for the $b$-th data pair in the mini-batch, and $B$ is the mini-batch size.
The error signal (derivative of the loss \eqref{MSE-loss} with respect to each phase vector $\boldsymbol \theta_n \in \mathbf{\Theta}$) is propagated back through the model to adjust the phases of the combining vectors \cite{LearningMachines} \cite{EffBackProp}, making up what is usually referred to as the backward pass or \textit{backpropagation}. This is formally expressed by the chain rule of differentiation:
\begin{equation}\label{chain_rule}
  \left(\frac{\partial \mathcal L}{\partial \boldsymbol{\theta_n}}\right)^T = \frac{\partial \mathcal L}{\partial g} \cdot \left(\frac{\partial g}{\partial \mathbf q}\right)^T \cdot \frac{\partial \mathbf q}{\partial \mathbf z} \cdot \frac{\partial \mathbf z}{\partial \boldsymbol{\theta_n}}.
\end{equation}
In mathematical terms, $\frac{\partial \mathcal L}{\partial \boldsymbol{\theta_n}}$ does not exist, for the factor $\frac{\partial \mathbf q}{\partial \mathbf z}$ does not satisfy the Cauchy-Riemann equations \cite{Haslinger2017}, meaning that $\mathbf q$ as a function of the complex vector ${\bf z}$ is not complex differentiable (\textit{holomorphic}). However, the issue could be resolved to enable backpropagation. The details of that and how the derivatives are computed are discussed in Appendices \ref{app:comp_diff} and \ref{app:comp_der}. Computing the partial derivative of the loss with respect to phase vector $\boldsymbol{\theta_n}$ allows the backward pass to modify the codebook $\boldsymbol \Theta$ and make it adaptive to the environment. The update equation generally depends on the solver used to carry out the training, e.g., Stochastic Gradient Descent (SGD) and ADAptive Moment estimation (ADAM) to name two, but in its simplest form, it could be given by
\begin{equation}\label{eq:update_rule}
  \boldsymbol{\theta_n}_{\mathsf{new}} = \boldsymbol{\theta_n}_{\mathsf{cur}} - \eta \cdot \frac{\partial \mathcal L}{\partial \boldsymbol{\theta_n}}
\end{equation}
where $\eta$ is the optimization step size, commonly known as the learning rate in machine learning, and $\boldsymbol{\theta_n}_{\mathsf{new}}$ and $\boldsymbol{\theta_n}_{\mathsf{cur}}$ are, respectively, the new and current $n$-th phase vector of the codebook.

\subsection{Learning Quantized Codebook}\label{sec:quan_cb}
Restriction on the resolution of the phase shifter is common in many mmWave implementations. This imposes limits on the number of phase vectors that could be realized by a system, giving rise to learning quantized codebooks. The proposed solution is actually capable of learning such codebooks. This could be achieved using a quantize-while-training approach, similar to that in \cite{TernQuant}\cite{DeepComp}. The training process, presented as forward and backward passes in Sections \ref{sup_arch} and \ref{sec:sup_learning}, respectively, is tweaked to incorporate a k-means quantizer. The quantizer is implemented right after updating the parameters of the network in \eqref{eq:update_rule}. It takes the phase-vector codebook and vectorize it $\boldsymbol{\tilde \Theta} = \left[ \boldsymbol{\theta}_1^T,\dots,\boldsymbol{\theta}_N^T \right]_{1\times NM}$, and then, it applies k-means on the elements of $\boldsymbol{\tilde \Theta}$. The returned cluster centroids, which are a set of scalars, define the new finite set of angles the phase shifters need to realize. The size of that set (number of centroids) is determined by the phase shifter resolution, $Q$ bits.

\section{Self-Supervised Machine Learning Solution}\label{self-sup_sol}

\begin{figure*}[t]
	\centering
	\includegraphics[width=\linewidth]{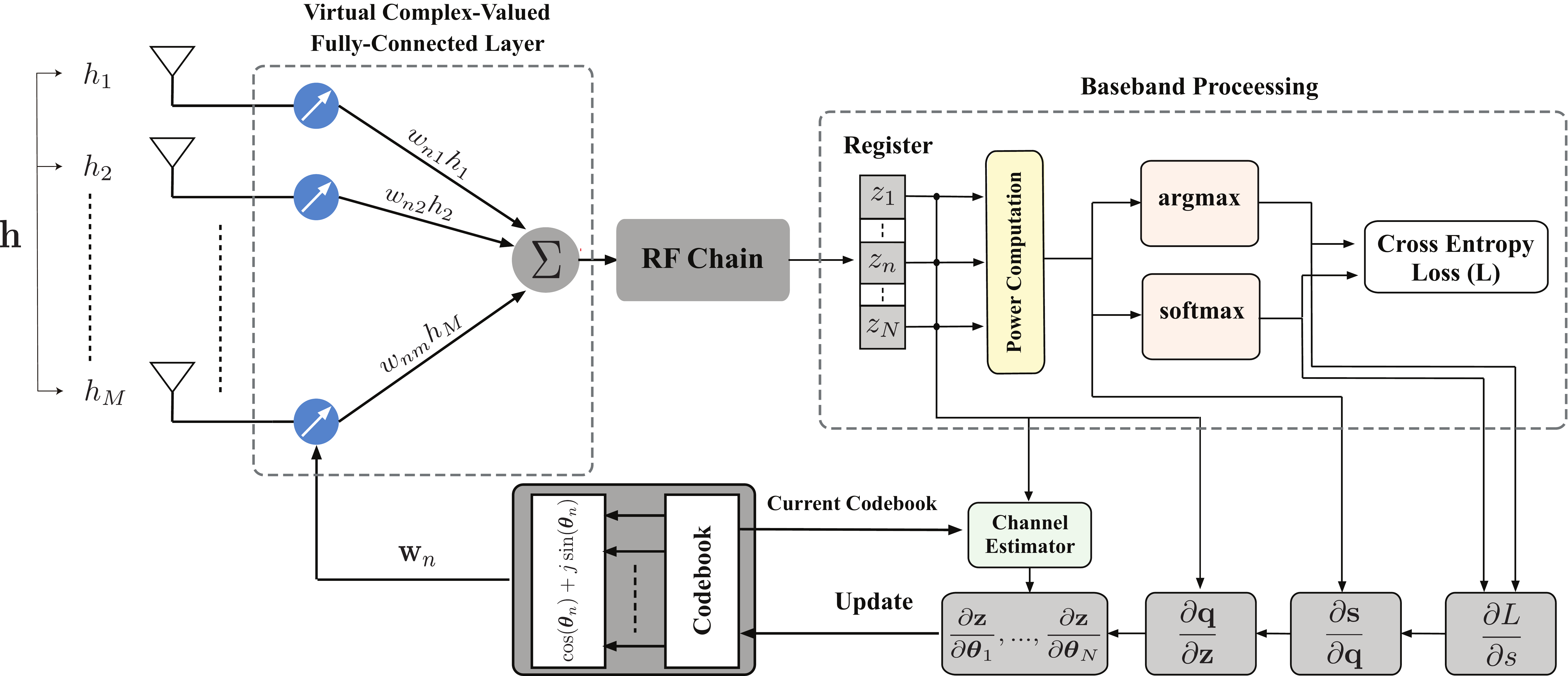}
	\caption{The proposed self-supervised framework as it is envisioned in practice. The solution is integrated into the different components of a mmWave base station with analog architecture. The phase shifters with the combiner all together form a fully-connected layer, and the rest of the layers are implemented into the base-band processing unit.}
	\label{selfsup_arch}
\end{figure*}

In this section, an alternative neural network architecture is proposed to perform the same codebook learning process without requiring accurate channel knowledge. The motivations for developing this model are two fold: (i) The existence of hardware impairments prevents accurate channel acquisition in mmWave systems as obtaining them could be a difficult process that requires very large training overhead \cite{Alkhateeb2014}, and (ii) the need for channel information implies that the codebook learning process has to be performed \textit{offline}, which may not be favorable for swift adaptability. To address these problems, we propose  a novel \textit{self-supervised} learning solution. This solution, as the name suggests, works in a self-sufficient fashion instead of requiring the supply of a desired response for every training channel.

\subsection{Self-supervision via Clustering}
Before diving into the details of the new proposed architecture, it is helpful to first illustrate the basic idea of this design. The motivations for this new model are rooted in the lack of desired responses and the need for online learning. As a result, the model should only rely on itself to learn how to adjust the codebook beams such that the performance is improved. This is accomplished by tapping into an intrinsic feature the final codebook must have, channel space partitioning; as explained in Section \ref{sec:Prob}, the codebook has fixed size, and, therefore, each beamformer is ultimately expected to be optimized to serve a set of users in the environment. This is mathematically equivalent to partitioning $\boldsymbol{\mathcal H}$ into  subsets of channels
\begin{equation}
  \boldsymbol{\mathcal H} = \boldsymbol{\mathcal H}_1 \cup \boldsymbol{\mathcal H}_2 \cup \dots \cup \boldsymbol{\mathcal H}_N,
\end{equation}
where
\begin{equation}
  \boldsymbol{\mathcal H}_{n^\prime} \cap \boldsymbol{\mathcal H}_n = \emptyset, \ \forall n^{\prime} \neq n\ \text{and} \ n^{\prime},n\in \{1,\dots,N\}.
\end{equation}
From a machine learning perspective, this partitioning could be translated into channel \textit{clustering} where each beamformer is a cluster representative. 
Under this new view of the problem, the machine learning model generates its labels using the following strategy: For the received signal of an uplink pilot, it identifies the best beamforming vector in the current codebook, say $\mathbf w_n$ where $n\in \{1,\dots,N\}$. Then, it adjusts the direction of that beamformer such that it results in higher beamforming gain with the current channel. Therefore, when a similar channel form the same partition, say $\boldsymbol{\mathcal H}_n$, is experienced, $\mathbf w_n$ is expected to be the best beamformer again, increasing its chance to be the representative for $\boldsymbol{\mathcal H}_n$. The technical details on how this is done are presented in the following couple of subsections, in which the model components, forward pass, and backward pass are explored.  

\subsection{Model Architecture}\label{self-sup_arch}

\figref{selfsup_arch} presents a schematic of the proposed architecture as it is envisioned in a mmWave communication system. The following details  a forward pass through the different components of this architecture: 

\subsubsection{Complex-Valued Fully-Connected Layer}
similar to its supervised counterpart, the self-supervised network also adopts complex-valued fully-connected layer as its first layer. However, as integration into the communication system is in the core of this solution, the layer is implemented using the phase shifters, not a digital processor. As a result, it is referred to in \figref{selfsup_arch} as a \textit{virtual complex-valued fully-connected layer} (virtual layer for short). The function this layer implements is the same as that in Section \ref{subsub:comp_val_fc}, and as such, its output is also given by \eqref{fc_1}. The main difference between this layer and that in Section \ref{subsub:comp_val_fc} comes in the implementation of the matrix vector multiplication. The virtual layer performs it by requiring the user to send a sequence of pilots, each of which is received with a different beamformer.

\subsubsection{Register} the register buffers the received single of each beamformer until a full sweep across the codebook is completed. This temporary storage is essential as the following layers need to operate on the outputs of the virtual layers jointly.
  
\subsubsection{Power-computation layer}
once the system collects the whole outputs (all the beams in the codebook have been tried), those values are fed into the power-computation layer which calculates the beamforming gain for each beamformer using \eqref{powOut}. 
\subsubsection{Softmax and argmax} this layer is where the self-supervised solution really differs from the supervised one. Instead of having a max-pooling layer, the output of power-computation layer is fed into two different layers, a softmax and an argmax. The former is employed to convert each beamforming gain to a ``probability'', which indicates how likely a beamformer is the optimal one to receive the user's signal given the current channel. Formally, having \eqref{powOut} as input, the $n$-th element of the output probability vector of the softmax layer ${\bf s}= [s_1,\dots,s_N]^T$ can be expressed as
\begin{equation}\label{outSoftmax}
  s_n = \frac{e^{|z_n|^2}}{\sum_{n=1}^{N}e^{|z_n|^2}}.
\end{equation}

The argmax layer, on the other hand, outputs a one-hot vector $\mathbf{c} \in \{0,1\}^N$, of the same dimension as ${\bf s}$, with $1$ at the position where ${\bf s}$ attains its maximum value and with $0$ at all other positions. This one-hot vector ${\bf c}$ is the self-generated label. It declares the best beamforming vector the representative of the cluster, and along with the output of softmax, they help tweak this best beamformer to make sure it has higher beamforming gain than other beamformers when it receives a similar channel in the future. This is accomplished by implementing a cross-entropy loss function. The following subsection will elaborate more on that loss and its role.

\subsection{Learning Codebooks}\label{sec:selfsup_learn}

After a forward pass, the model must do backpropagation to improve its performance, i.e., learning better beamforming vectors. With the self-generated label and the probability vector, it is a matter of using that label to increase the probability of the currently selected beamformer.

\subsubsection{Loss function} 
The first step to do backpropagation is to define a loss function that captures the objective of the model. As stated above, the model aims at clustering the channels and having the beamforming vectors in the codebook as representatives of those clusters. This is attained by a cross-entropy loss function given as
\begin{equation}\label{loss}
  \mathcal{L}({\bf s}, {\bf c}) = -\sum_{n=1}^{N}c_n\log s_n,
\end{equation}
where ${\bf s}$ is the output of the softmax layer and ${\bf c}$ is the one-hot vector generated by argmax layer. This loss function makes the one-hot vector a target probability distribution for the model, and hence, it is not adjustable; the value of $\mathcal L$ must only be minimized by pushing the softmax distribution $\mathbf s$ to be as close to $\mathbf c$ as possible.

\subsubsection{Backpropagation}
The error signal is generated by differentiating the loss \eqref{loss} with respect to each phase vector $\boldsymbol{\theta_n}\in\boldsymbol{\Theta}$. This error is backpropagated through the network to adjust the phases of all the beamforming vectors \cite{LearningMachines}\cite{EffBackProp} using the chain rule as follows
\begin{equation}\label{chain_rule_2}
  \left(\frac{\partial\mathcal{L}}{\partial\boldsymbol{\theta}_n}\right)^T
  = \left(\frac{\partial\mathcal{L}}{\partial{\bf s}}\right)^T \cdot \frac{\partial{\bf s}}{\partial{\bf q}} \cdot
  \frac{\partial{\bf q}}{\partial{\bf z}} \cdot \frac{\partial{\bf z}}{\partial\boldsymbol{\theta}_n},
\end{equation}
and \eqref{eq:update_rule} is used to update the phase vectors of the codebook. The implementation of this chain of derivatives is illustrated in \figref{selfsup_arch}.
There are two issues with the error signal in \eqref{chain_rule_2}. The first is similar to that issue encountered with the supervised model; ${\bf q}$ as a function of ${\bf z}$ is not \textit{complex differentiable} or \textit{holomorphic}, which implies that $\frac{\partial{\bf q}}{\partial{\bf z}}$ is not defined. The same argument developed for \eqref{chain_rule} and presented in Appendices \ref{app:comp_diff} and \ref{app:comp_der} will be used here to obtain that partial. The second issue comes from the partial $\frac{\partial{\bf z}}{\partial\boldsymbol{\theta}_n}$. Referring to \eqref{fc_1}, it is clear that computing $\frac{\partial{\bf z}}{\partial\boldsymbol{\theta}_n}$ requires channel information, which is not explicitly available in this case. This is sidestepped with the help of a simple channel estimator described in the following subsection.

\subsubsection{Channel Estimator}
In order to complete the backpropagation of the error signal, the content of the register in \figref{selfsup_arch} is also fed to a channel estimator. This estimator uses the received signals along with the currently available beamforming codebook to reconstruct a \textit{rough} estimate of the channel. Based on \eqref{fc_1}, we notice that the output $z_n$ of each combiner ${\bf w}_n$ is essentially the projection of the channel ${\bf h}$ onto the subspace spanned by the combiner ${\bf w}_n$. Thus, we estimate a rough version of the channel through
\begin{equation}\label{estCh}
  \widehat{{\bf h}} = \left(\mathbf W^H\right)^\dagger{\bf z}.
\end{equation}
This approach does not result in an accurate estimate of the channel, yet it helps the learning process as shown in Appendix \ref{app:comp_der}.

\section{Practicality of Proposed Solutions}\label{sec:pract_impl}
Both proposed solutions are developed with practicality in mind. They are both geared towards handling different challenges commonly faced in designing mmWave beam codebooks, especially with fully-analog architectures. However, that does not mean they operate in the same way. They approach the codebook learning problem from different angles, as briefly discussed below. 

The supervised learning solution relies on explicit channel knowledge and follows a \textit{transparent} leaning approach. It requires the mmWave system to operate with some common environment-independent codebook, like the DFT codebook, and during its operation it collects channel information from the surroundings. Such information is used to construct the training dataset ($\mathcal S_t$) as described in Section \ref{sec:sup_learning}. Once a dataset is available, the central unit trains the model in the background, and upon the completion of the training phase, the new environment-aware codebook is directly plugged into the system. This method decouples the communication operation from the codebook learning process, and allows the system to function normally until a better codebook is learned. Its main drawbacks, however, are the requirement of accurate (or good quality) channel estimates to construct the training dataset, and the relatively lengthy  offline learning process. 

The self-supervised solution, in contrast, trades explicit and accurate channel knowledge for faster training and adaptation. The need for accurate channel estimates in itself is a burden to the mmWave system, especially when hardware impairments are factored in. Hence, the self-supervised solution is designed to transcend that need. As shown in Section \ref{sec:selfsup_learn}, the model is implemented as an integral component of the mmWave system and does not run in the background. The learning instead is performed online while the system is operating. This provides a more adaptable and faster training in terms of implementation. However, this adaptability comes with its own shortcomings. The first one is a subtle degradation in the quality of the learned codebook compared to that of the supervised solution (as will be discussed in Section \ref{sec:results}). It is a direct consequence of implementing a simple yet noisy channel estimator. The other issue is an unstable communication performance at the beginning of the learning process. Different to the transparent nature of the supervised solution, the self-supervised solution learns on the job, and as a results the codebook itself evolves over time.

\begin{figure}[t]
	\centering
	\subfigure[LOS Scenario]{ \includegraphics[width=0.45\linewidth]{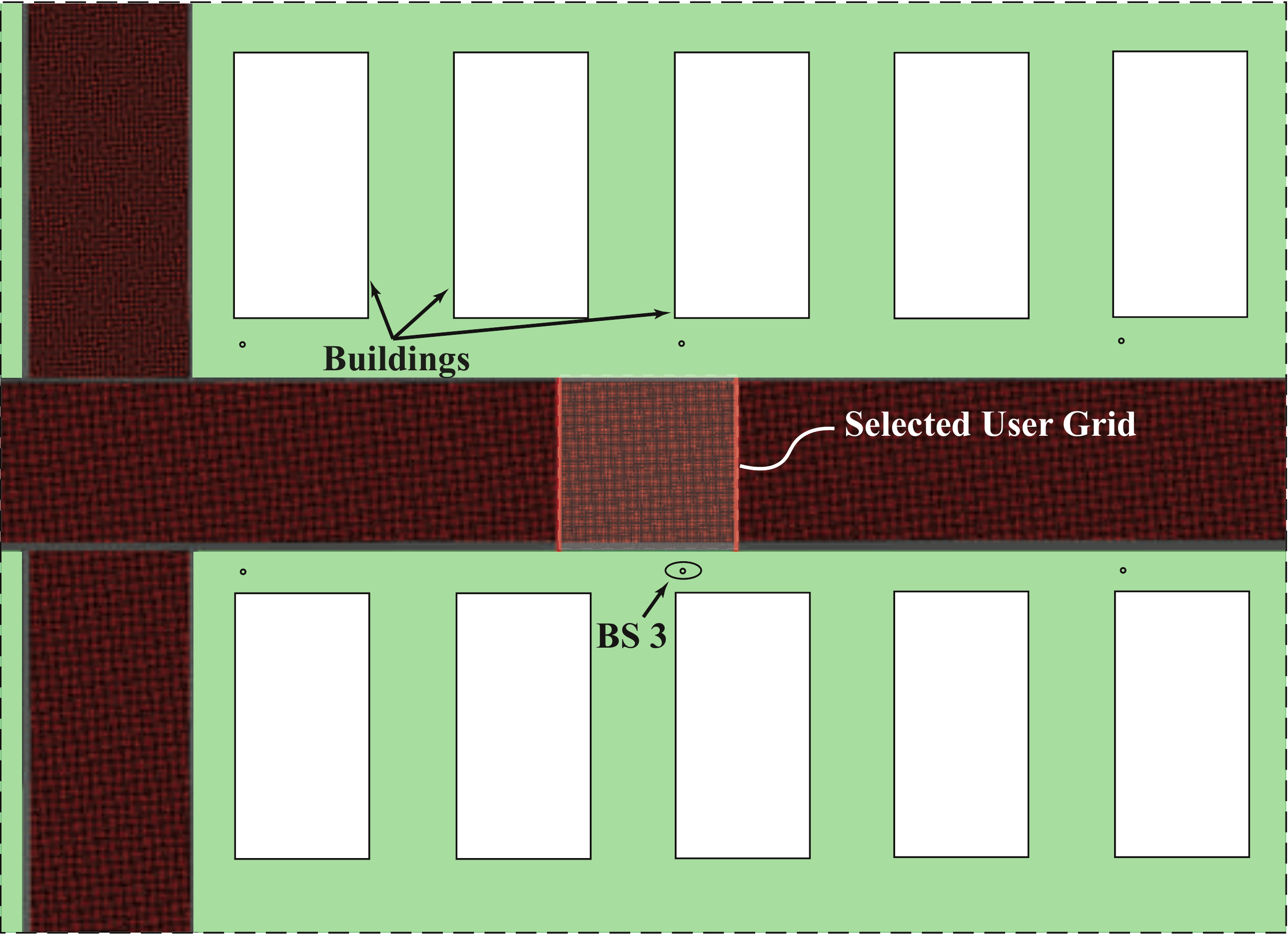}\label{fig:sce_los} }
	\subfigure[NLOS Scenario]{ \includegraphics[width=0.45\linewidth]{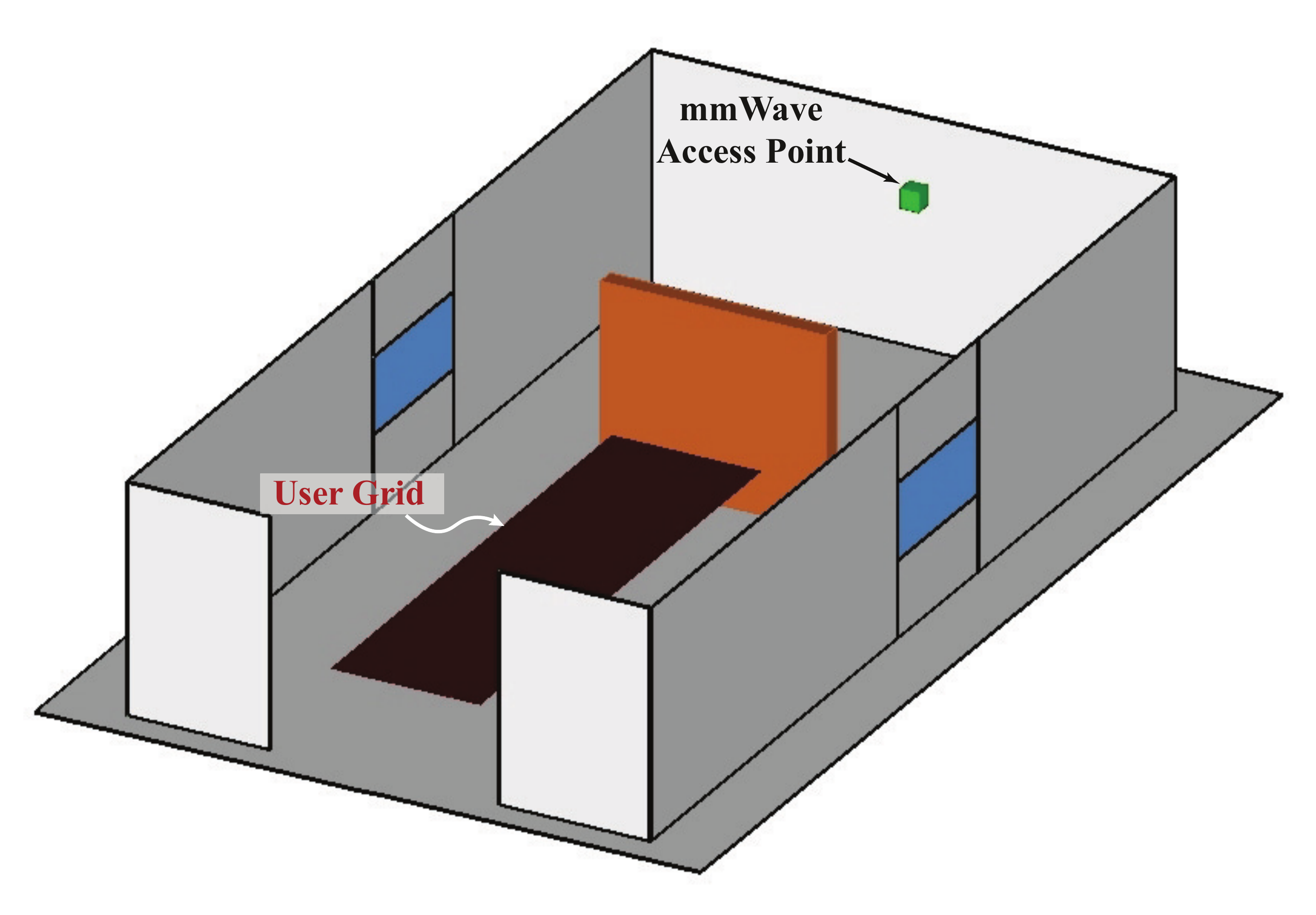}\label{fig:sce_nlos} }
	\caption{Two perspective views of the considered communication scenarios. (a) shows the LOS scenario. It is chosen to be outdoor since the likelihood of LOS connection is higher there. (b) shows the NLOS scenario. Similar to (a), this scenario has been chosen for the high likelihood of having NLOS users indoors.}
	\label{scenarios}
\end{figure}

\section{Experimental Setup and Model Training} \label{sec:exp_setup}
In order to evaluate the performance of the proposed codebook learning solutions, two communication scenarios are considered. They are designed to represent two different communication settings. The first has all users experiencing LOS connection with the basestation while the other has them experiencing NLOS connection. The following two sections provide more details on the scenarios and the training and testing processes.

\subsection{Communication Scenarios and Datasets}\label{sec:scen_data}
Two communication scenarios are used for performance evaluation. The first one is, as mentioned earlier, a LOS scenario, see \figref{scenarios}-(a). It is an outdoor scene where all users have LOS connection with the mmWave base station. The second scenario, on the other hand, is chosen to be an indoor NLOS scenario where all users have NLOS connection with the mmWave base station. Both scenarios are for an operating frequency of 28 GHz, and both are part of the DeepMIMO dataset \cite{DeepMIMO}. 
Using the data-generation script of DeepMIMO, two sets of channels, namely $\mathcal S^\rm{LOS}$ and $\mathcal S^\rm{NLOS}$, are generated, one for each scenario. Table \ref{param} shows the data-generation hyper-parameters. For the supervised solution, both sets undergo processing to generate the labels and create two sets of pairs as described in Section \ref{sec:back_trn}. The new datasets are henceforth referred to as $\mathcal S^\rm{LOS}_{t_1}$ and $\mathcal S^\rm{NLOS}_{t_1}$. For the self-supervised solution, on the other hand, labels are not needed, and, therefore, the two sets $\mathcal S^\rm{LOS}$ and $\mathcal S^\rm{NLOS}$ are used as they are. For the sake of convenience, these two sets will be re-named $\mathcal S^\rm{LOS}_{t_2}$ and $\mathcal S^\rm{NLOS}_{t_2}$.

\begin{table}[t]
\caption{Hyper-parameters for channel generation}
\centering
\begin{tabular}{|c | c | c|}
\hline
	Parameter & \multicolumn{2}{c|}{value} \\
	\hline\hline
    Name of scenario & O1\textunderscore28 & I2\textunderscore28B \\
    \hline
    Active BS & 3 & 1 \\
    \hline
    Active users & 800 to 1200 & 1 to 700 \\
    \hline
    Number of antennas (x, y, z)  & (1, 64, 1) & (64, 1, 1) \\
    \hline
    System BW & 0.2 GHz & 0.2 GHz\\
    \hline
    Antenna spacing & 0.5 & 0.5 \\
    \hline
    Number of OFDM sub-carriers & 1 & 1 \\
    \hline
    OFDM sampling factor & 1 & 1 \\
    \hline
    OFDM limit & 1 & 1 \\
    \hline
\end{tabular}
\label{param}
\end{table}

\subsection{Model Training}
The two models are trained and tested on their datasets introduced in the earlier section, Section \ref{sec:scen_data}. The training of both solutions follow the same strategy. It starts by data pre-processing. The channels in each dataset are normalized to improve the training experience \cite{EffBackProp}, which is a very common practice in machine learning. As in \cite{DetChPred}\cite{LIS},\cite{OneBitADC} and \cite{CoordBeamForm}, the channel normalization using the maximum absolute value in the training dataset helps the network undergo a stable and efficient training. Formally, the normalization factor is found as follows
\begin{equation}
  \Delta = \underset{\mathbf h\in S}{\max} \ | h_{m,u} |^2
\end{equation}
where $h_{m,u}\in \mathbb C$ is the $m$th element in the channel vector of the $u$th user, and $\mathcal S \in \{\mathcal S^\rm{LOS}_{t_1},\mathcal S^\rm{NLOS}_{t_1}, \allowbreak \mathcal S^\rm{LOS}_{t_2},\mathcal S^\rm{NLOS}_{t_2}\}$.
Using the normalized channels, each solution is, then, trained on portion of the samples of the dataset and validated on the rest. The data split percentage between training and testing along with other training hyper-parameters are listed in Table \ref{train_hyp}. Example model-training scripts of the developed codebook learning solutions are available in \cite{myGithub} and \cite{Yu_github}.

\begin{table}[t]
\caption{Hyper-parameters for model training}
\centering
\begin{tabular}{|c | c | c|}
\hline
	Parameter & \multicolumn{2}{c|}{value} \\
	\hline\hline
    Solution & Supervised & Self-supervised \\
    \hline
    Batch size & 500 & 500 \\
    \hline
    Learning rate & 0.1 & 0.1 \\
    \hline
    Epoch number & 5 & 5 \\
    \hline
    Data split (training-testing) & 70\%-30\% & 70\%-30\% \\
    \hline
\end{tabular}
\label{train_hyp}
\end{table}

\section{Simulation Results} \label{sec:results}

In this section, we evaluate the performance of the proposed solutions using  the scenarios described in \sref{sec:exp_setup}. The numerical results show that our proposed models can adapt to both different environments and user distributions as well as imperfect array manufactures, meaning that the proposed codebook learning approaches  are  aware of the deployment and the hardware.

\subsection{Simulation Results for the Supervised Solution}

\begin{figure}[t]
	\centering
	\subfigure[LOS scenario]{\includegraphics[width=.45\columnwidth]{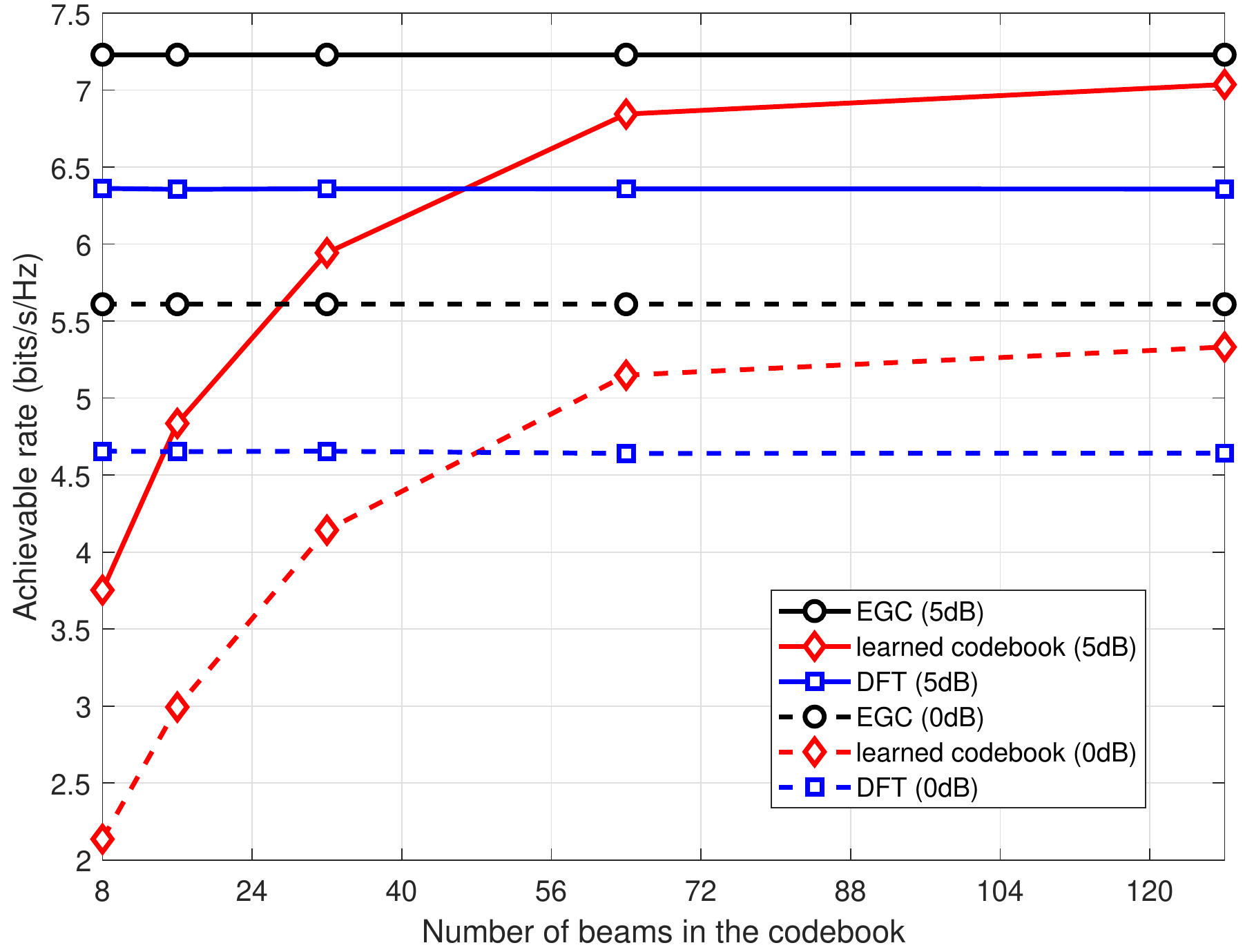}\label{sup_noise_0dB_5dB_LOS}}
	\subfigure[NLOS scenario]{\includegraphics[width=.45\columnwidth]{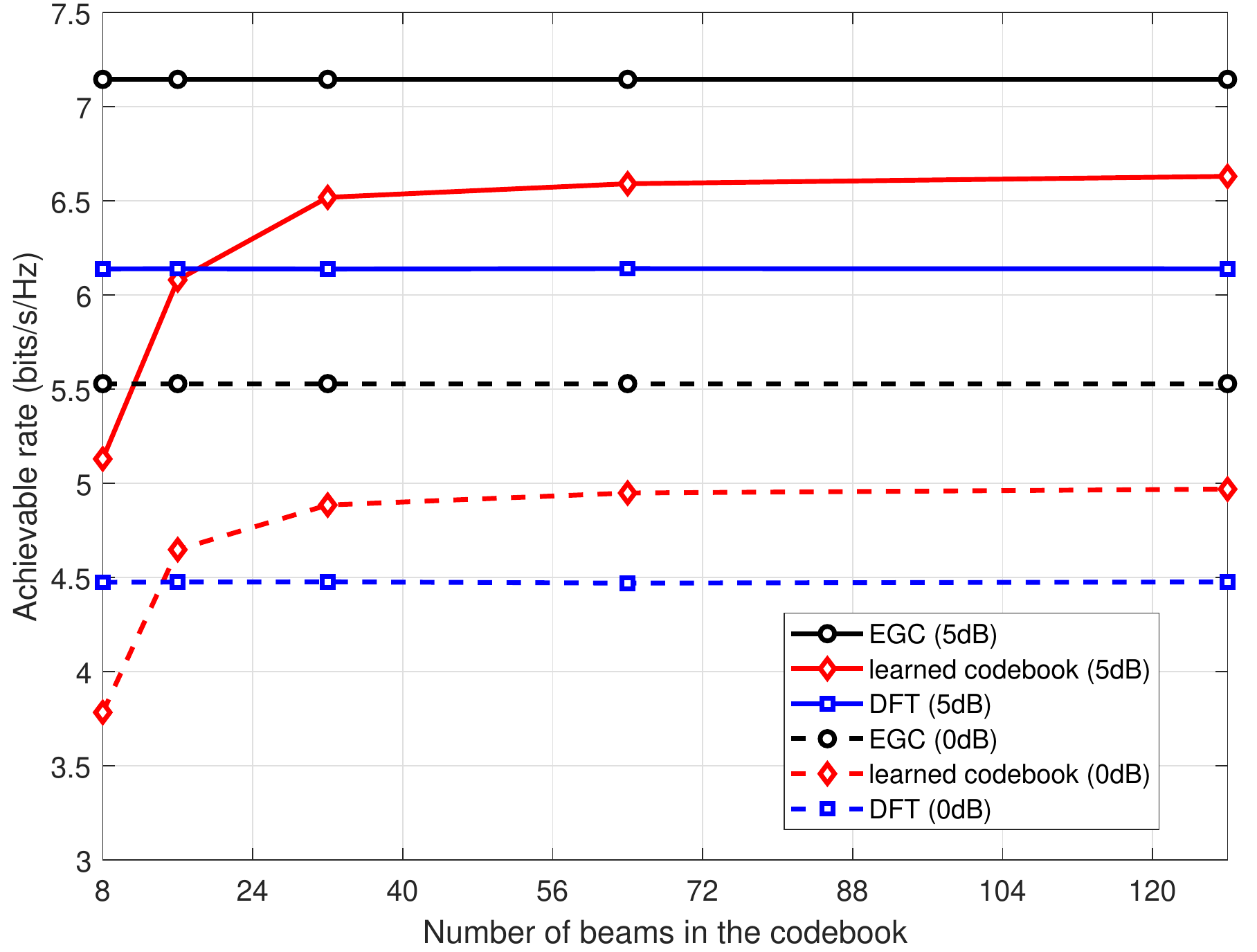}\label{sup_noise_0dB_5dB_NLOS}}
	\caption{The achievable rate versus the number of beams of the codebook using the supervised solution in: (a) LOS scenario and (b) NLOS scenario. It shows the performance under two receive SNRs, 0 and 5 dB.}	
\end{figure}

The performance of the proposed supervised solution is studied first in a LOS setting. \figref{sup_noise_0dB_5dB_LOS} shows the achievable rate versus the codebook size under 0 dB and 5 dB SNRs. The learned codebook exhibits interesting behavior compared to a 64-beam DFT codebook and an EGC receiver. With half the number of beams of a DFT codebook, the learned codebook achieves more than $80\%$ of the rate that the DFT achieves. This is very important and interesting as smaller codebook size means less beam training overhead. 
Further, this figure shows that when the learned codebook is allowed to have the same number of beams as the DFT codebook,  the performance of the proposed solution clearly surpasses that of the  DFT beam-steering codebooks. This is quite intriguing as, typically, a DFT codebook performs very well in a LOS setting. Generally, the supervised solution can produce a codebook that gets closer to the upper bound (EGC receiver) than a DFT codebook could, which is an immediate result of its adaptability.

The solution is also evaluated in a NLOS setting, which is expected to be more interesting and challenging; in a NLOS scenario, there is usually no single dominant path from a user to the base station, but there are multiple almost equally-dominant paths reflecting off of some scatterers. Therefore, to achieve good performance, a codebook should be able to capture as much of those dominant paths as possible such that the average received SNR after beamforming is increased. Similar to the LOS case, \figref{sup_noise_0dB_5dB_NLOS} depicts the achievable rate of the proposed solution versus the number of beams. \textbf{What is interesting here is how the learned codebook outperforms the 64-beam DFT codebook with way less number of beams, only 16 beams are enough to match the DFT performance}---the reason behind that will be discussed in the following paragraph. As the number of learned beams increases, its performance edges closer to the upper bound, achieving almost $90\%$ of the upper bound with 64 beams.

\begin{figure}[t]
	\centering
	\subfigure[LOS]{\includegraphics[width=.24\linewidth]{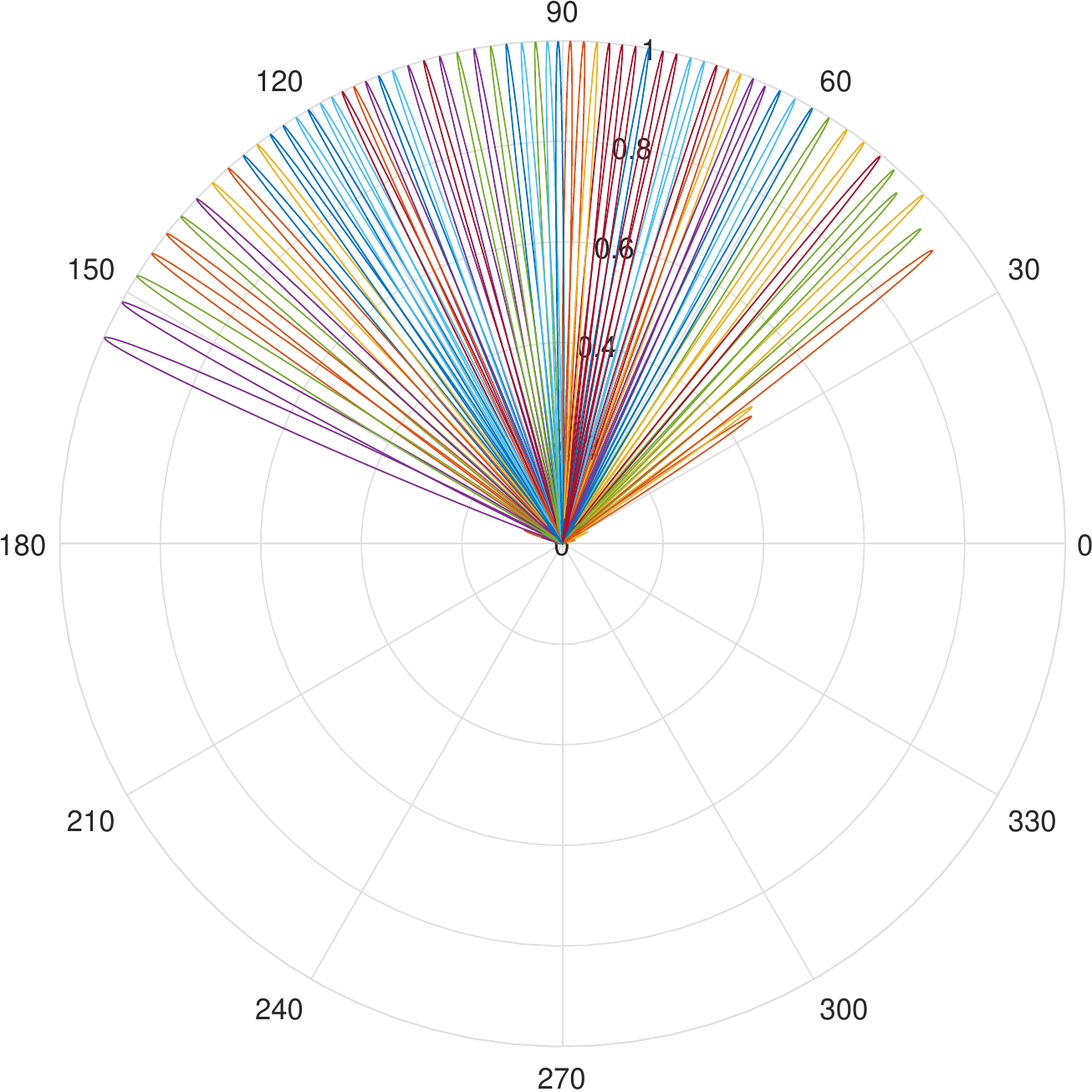}\label{fig:cb_los}}
	\subfigure[NLOS]{\includegraphics[width=.24\linewidth]{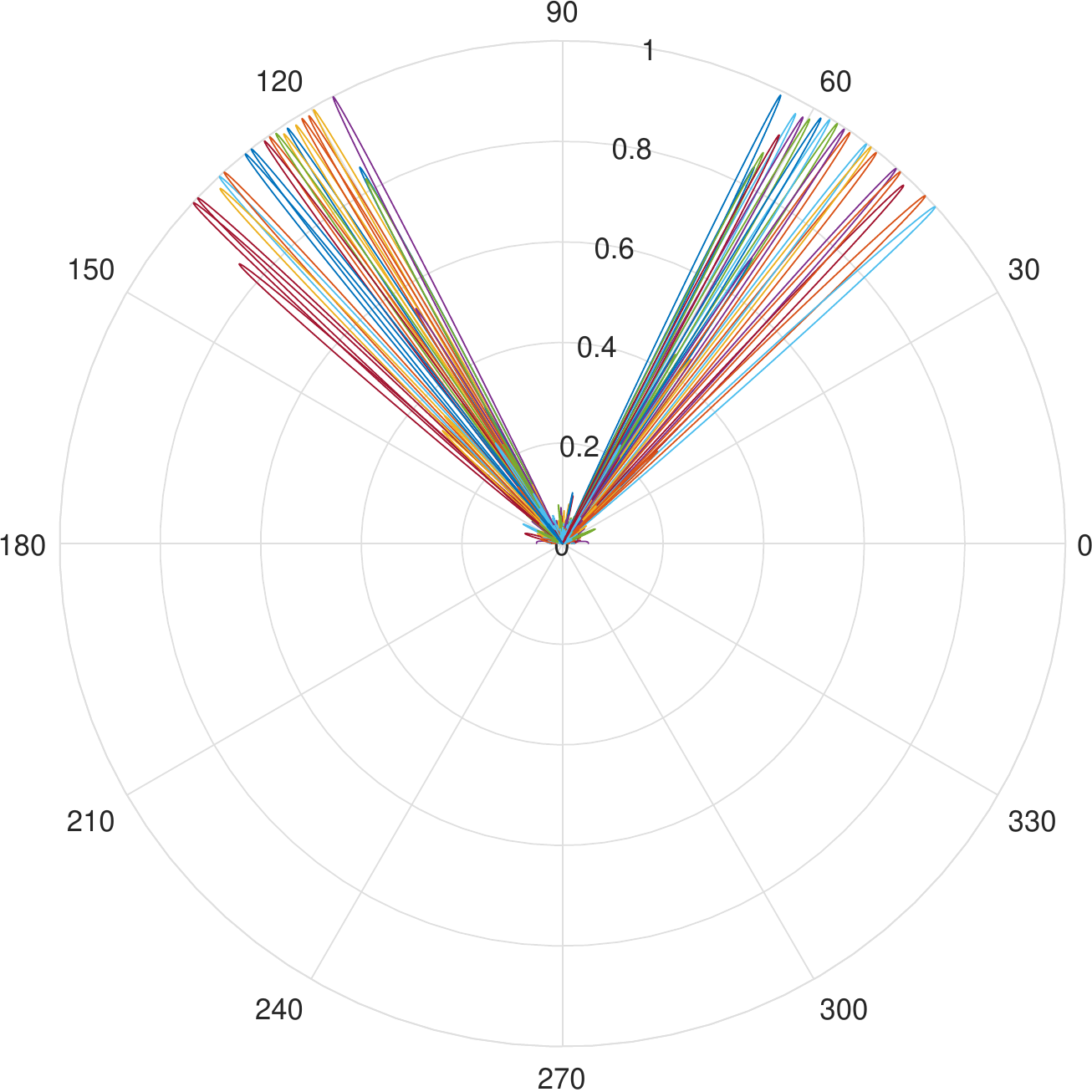}\label{fig:cb_nlos}}
	\subfigure[Single-beam, NLOS]{\includegraphics[width=.24\linewidth]{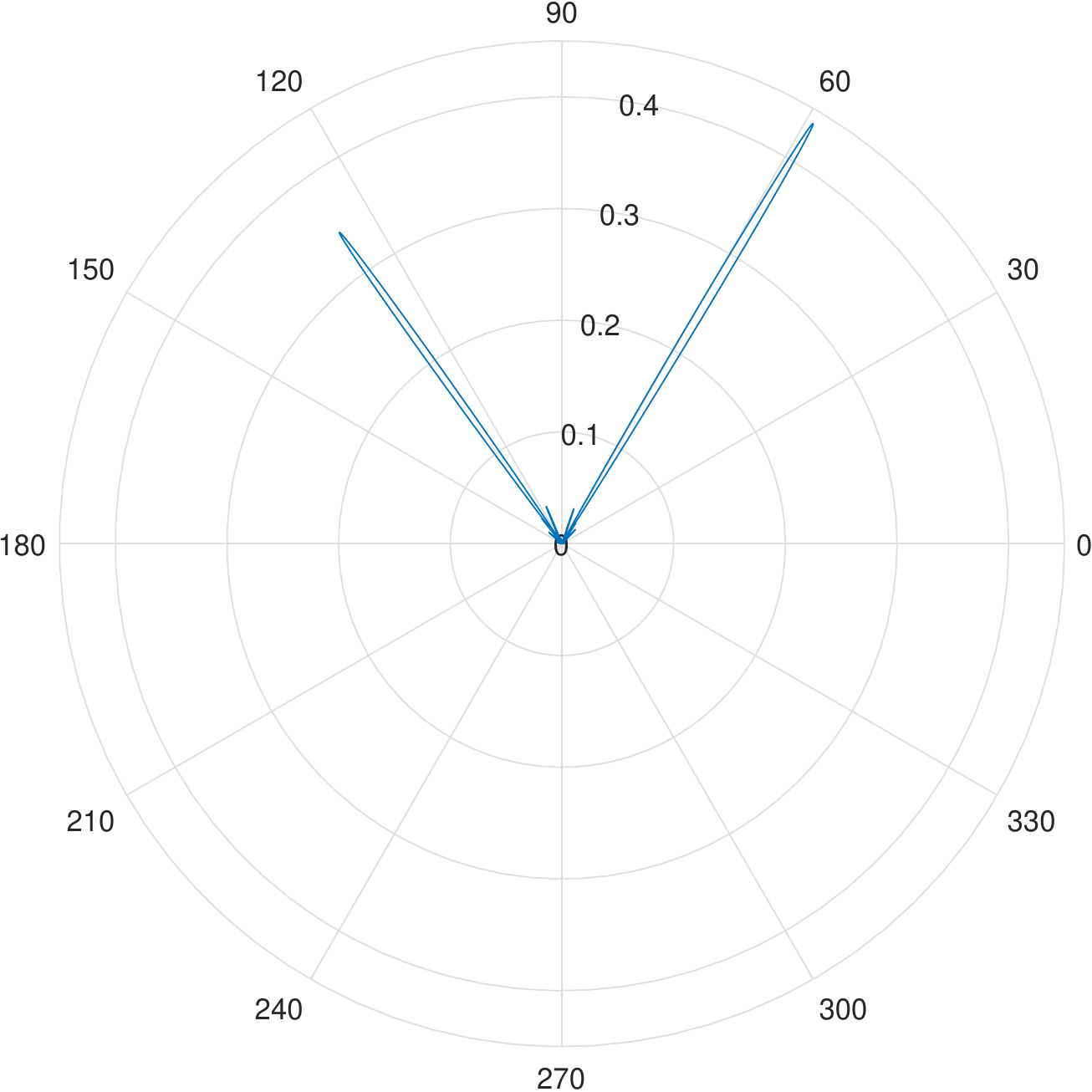}\label{fig:nlos_b1}}
	\subfigure[Single-beam, NLOS]{\includegraphics[width=.24\linewidth]{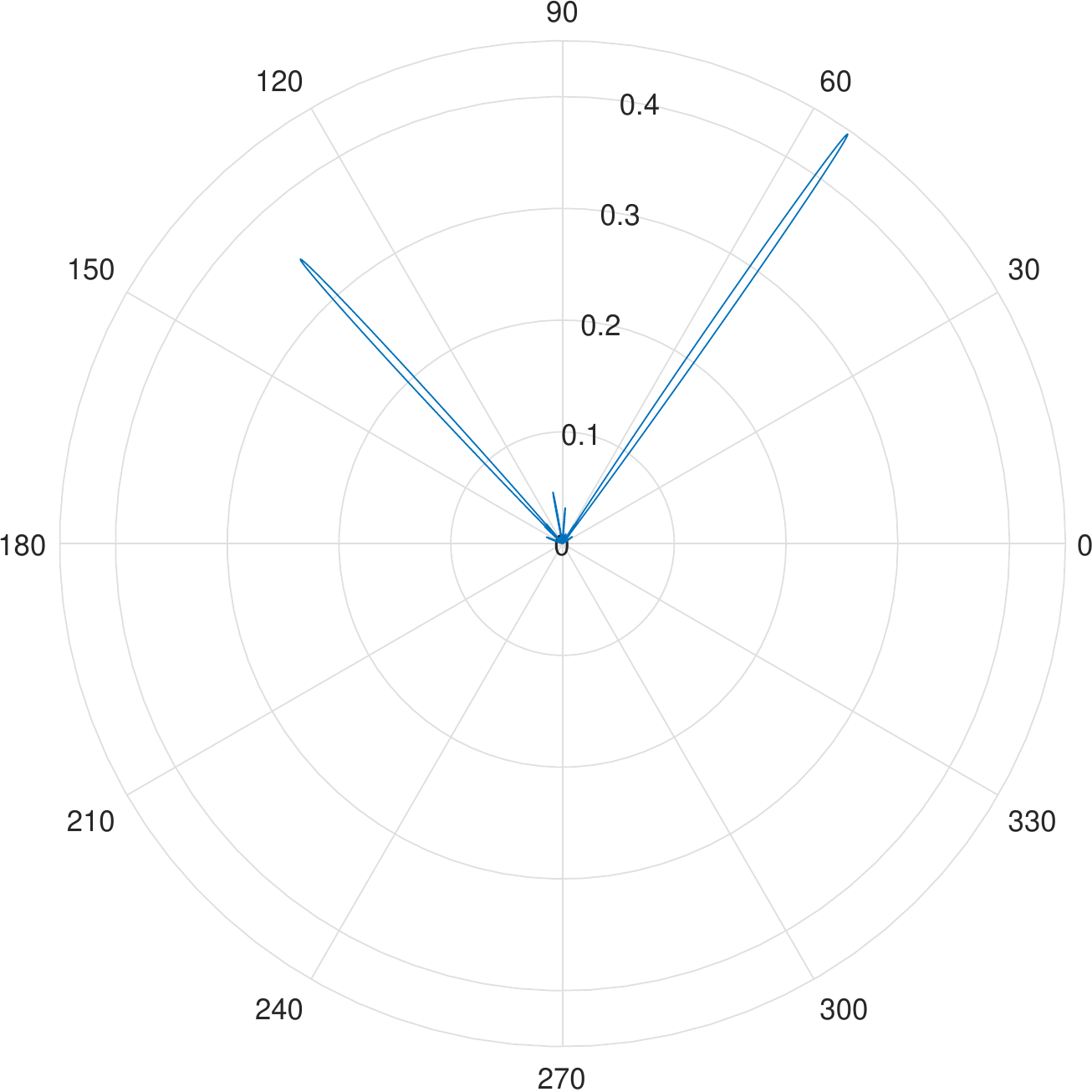}\label{fig:nlos_b2}}
	\caption{Beam patterns for the learned codebook using the supervised solution. (a) shows the codebook learned for the LOS scenario while (b) shows that learned for the NLOS scenario. Two beams from the 64-beam NLOS codebook are singled out in (c) and (d). They clearly show that the proposed solution is capable of learning multi-lobe beams.}
	\label{sup_64beams}
\end{figure}

To develop a deeper understanding of the performance of the proposed solution and verify its capability of learning beams that adapt to the surrounding environment and user distributions, we plot the the resulting beam patterns in \figref{sup_64beams}. More specifically, this figure shows different beam patterns of two different 64-beam codebooks learned in LOS and NLOS settings. The patterns in \figref{fig:cb_los} are for the LOS case, and they explain the improvement the learned 64-beam codebook experiences compared to the DFT codebook. Similar to the DFT codebook, all the learned beams are directive and have single-lobe, yet they do not spread across the whole azimuth plane like the DFT beams do. Their spread, instead, follows the user distribution in the scenario, the red rectangle drawn in \figref{fig:sce_los}. This makes each beam in the codebook tuned to serve a certain group of users and none of the beams is ``wasted'' by any means. In the NLOS setting, \figref{fig:cb_nlos} shows how the solution captures the different NLOS paths in the environment; the codebook is almost evenly split between the two major scatterers, the two side walls of the room. As a matter of fact, \textbf{looking at \figref{fig:nlos_b1} and \figref{fig:nlos_b2} reveals that the learned beams are not exclusively single-lobe, as some beams have multiple lobes that adapt to the main scatterers in the room}. This is a quite important property for a NLOS beam codebook, and it is evident in the codebook performance in \figref{sup_noise_0dB_5dB_NLOS}; it explains the clear gap in performance between the learned and DFT codebooks.

\begin{figure}[t]
	\centering
	\includegraphics[width=0.5\linewidth]{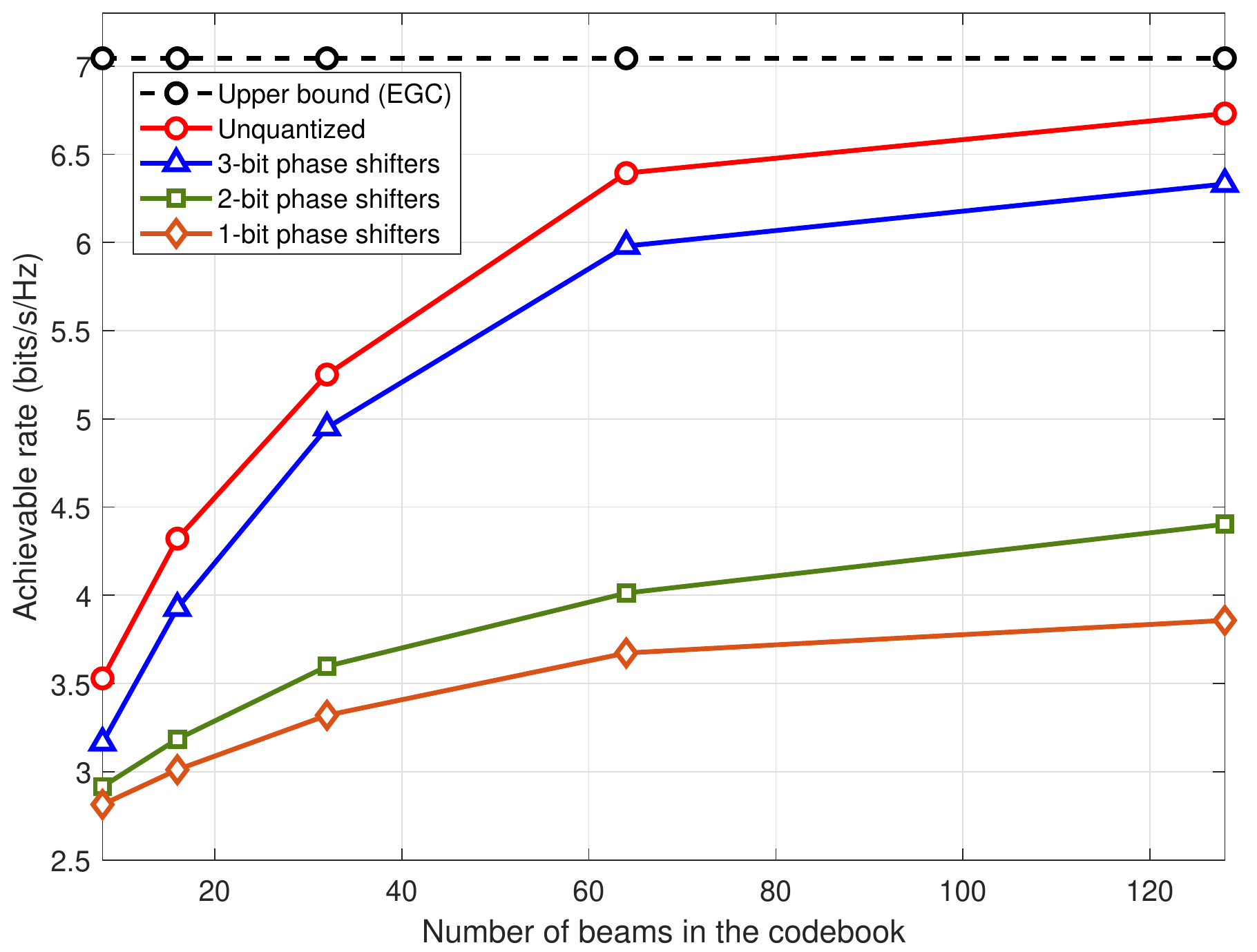}
	\caption{The achievable rate versus the size of the codebook in LOS setting. The figure shows the performance under different choices of quantized phase shifter.}
	\label{fig:quant_cb}
\end{figure}

To account for quantized phase shifters, the quantization method introduced in Section \ref{sec:quan_cb} is applied during the training of the supervised model to obtain a quantized codebook. \figref{fig:quant_cb} shows the performance of the learned codebooks with different phase quantization levels (i.e., number of bits). Despite its simplicity, this quantization approach can achieve over $80\%$ of the performance of the full-resolution phase shifters using only 3-bit phase shifters. This performance is consistent across all codebook sizes. This is very important and interesting for cases where the resolution of the analog phase shifters is limited.  


\subsection{Simulation Results for the Self-supervised Solution}
\begin{figure}[t]
	\centering
	\subfigure[]{\includegraphics[width=.45\columnwidth]{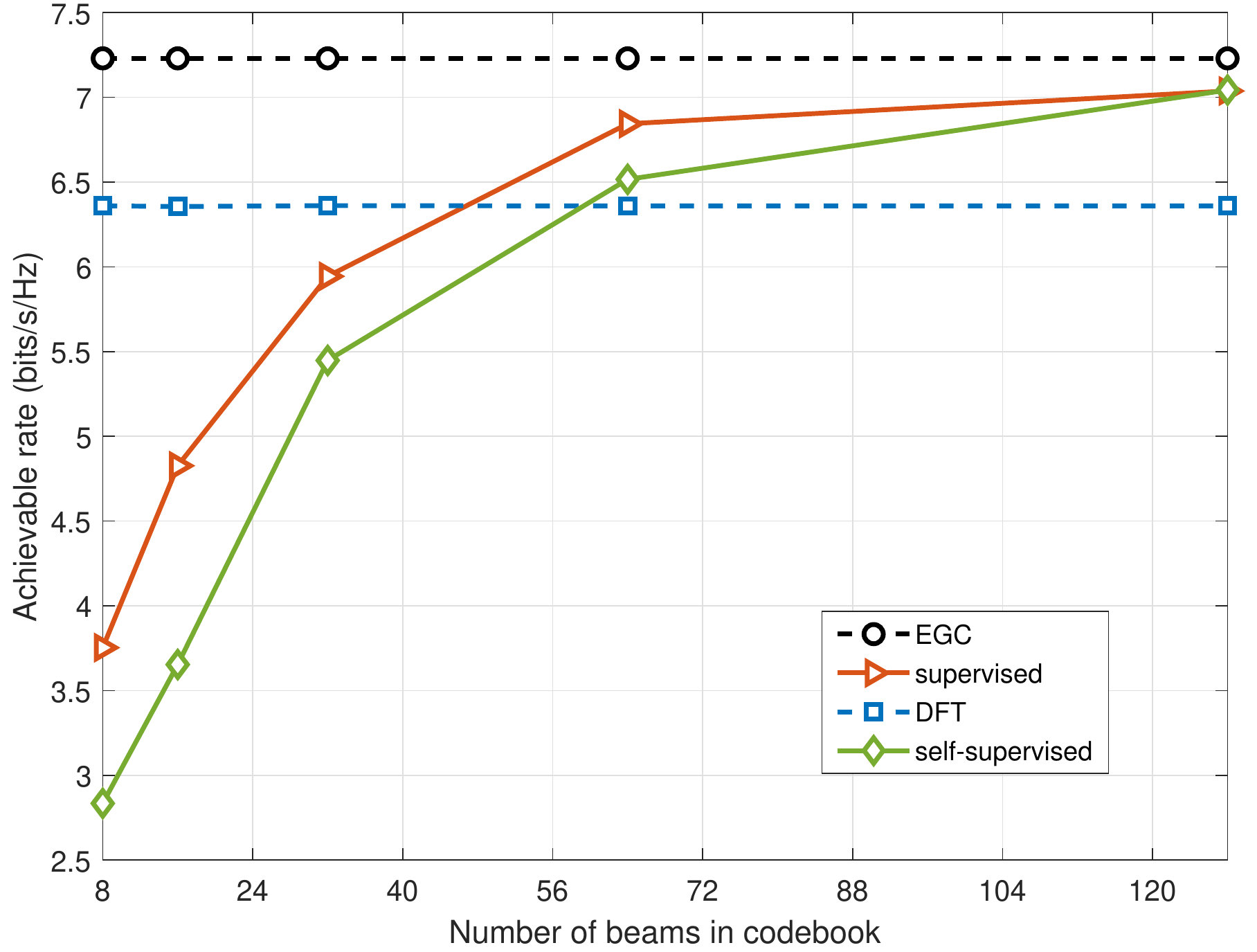}\label{sup_online_noise_5dB_LOS}}
	\subfigure[]{\includegraphics[width=.45\columnwidth]{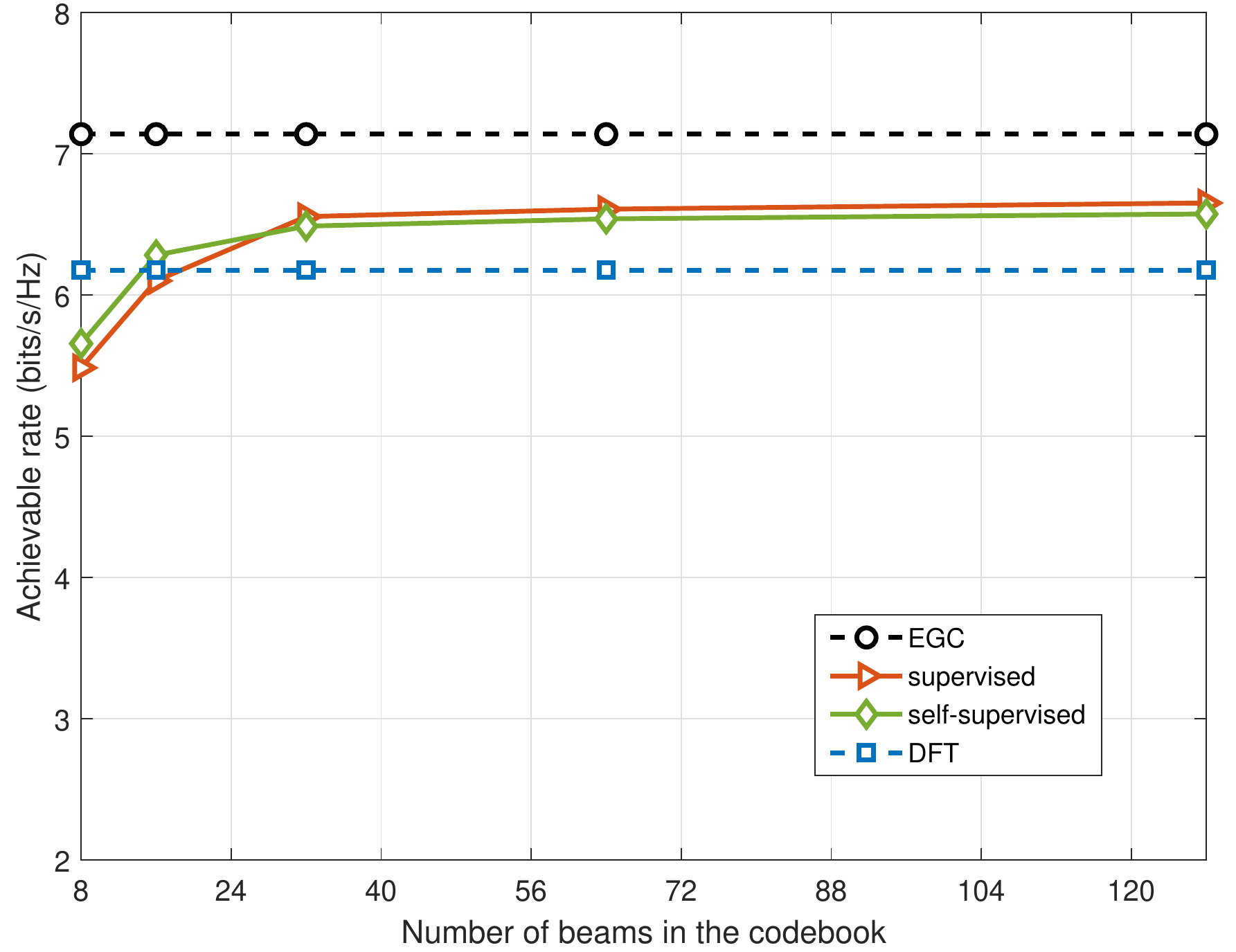}\label{sup_online_noise_5dB_NLOS}}
	\caption{The achievable rate vs. number of beams in the codebook with supervised and self-supervised learning solutions in: (a) a LOS scenario and (b) NLOS scenario. Both figures have the results for 5 dB receive SNR.}
\end{figure}

The performance of the self-supervised solution is benchmarked to that of the supervised one in both LOS and NLOS settings.
 \figref{sup_online_noise_5dB_LOS} plots the achievable rate versus the number of beams for both solutions in a LOS setting and under 5 dB SNR. This figure shows that the self-supervised codebook has a relatively similar performance to that of the supervised solution. As \figref{sup_online_noise_5dB_LOS} demonstrates, the self-supervised approach achieves over $90\%$ and $95\%$ of the achievable rates obtained by the supervised solution using  32 beams and 64 beams. In addition, the gap between the two solutions shrinks as more beams are learned. This comparable performance could be immediately extended to the NLOS case, as depicted in \figref{sup_online_noise_5dB_NLOS} which plots the achievable rate of the different approached versus codebook size under 5 dB SNR. 
\textbf{These results are very intriguing and promising as the self-supervised solution achieves this performance without the explicit channel knowledge.} This is an important property as stated in the beginning of Section \ref{self-sup_sol}. Channel estimation in fully-analog mmWave architectures is a considerable burden, and when hardware impairments are factored in, that burden amplifies. Hence, shedding light on the importance of that property make up the core of the following section.

\subsection{Performance Evaluation Under Hardware Impairments}

\begin{figure}[t]
	\centering
	\subfigure[]{ \includegraphics[width=.45\columnwidth]{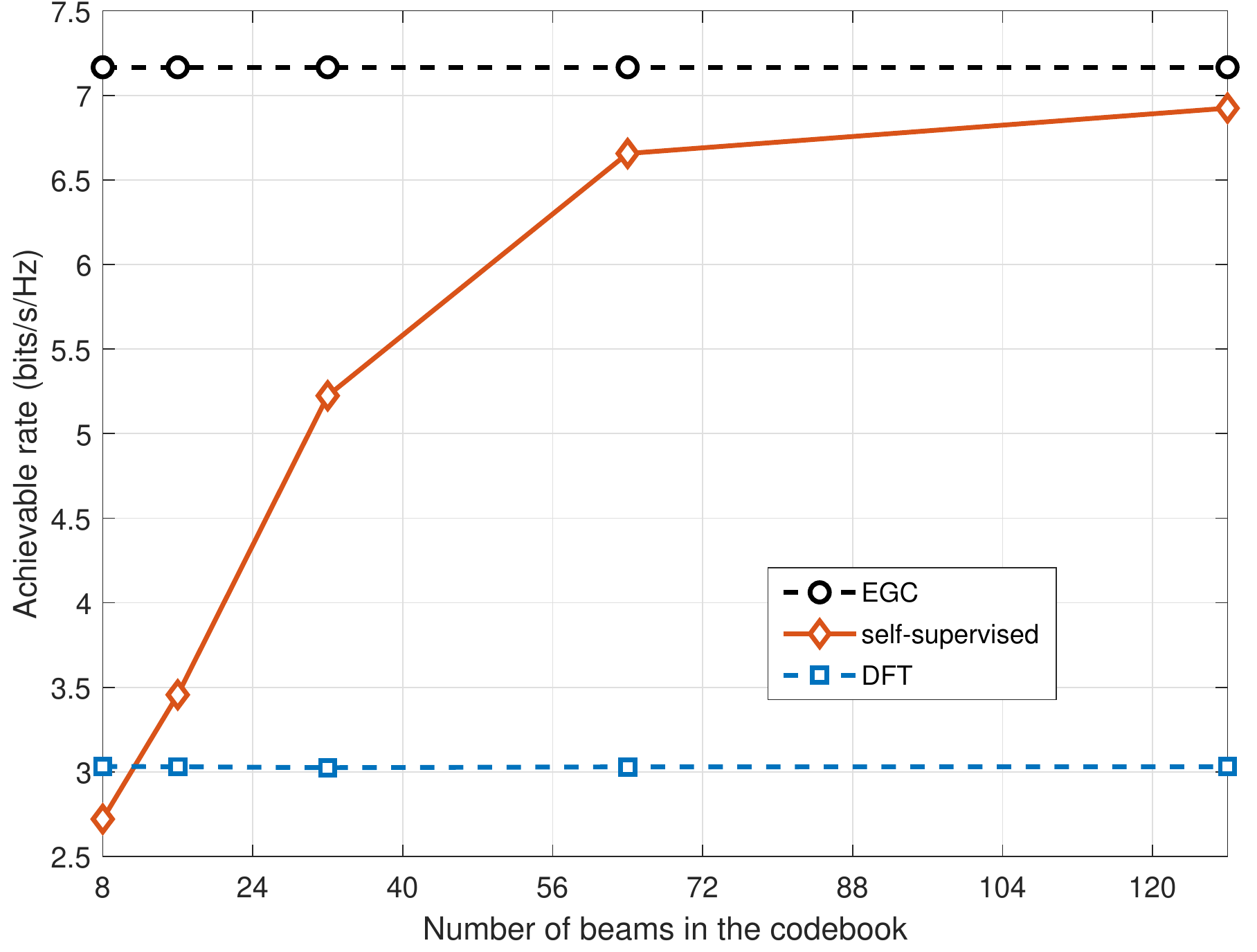}\label{sup_online_impairment_noise_5dB_LOS}}
	\subfigure[]{\includegraphics[width=.45\columnwidth]{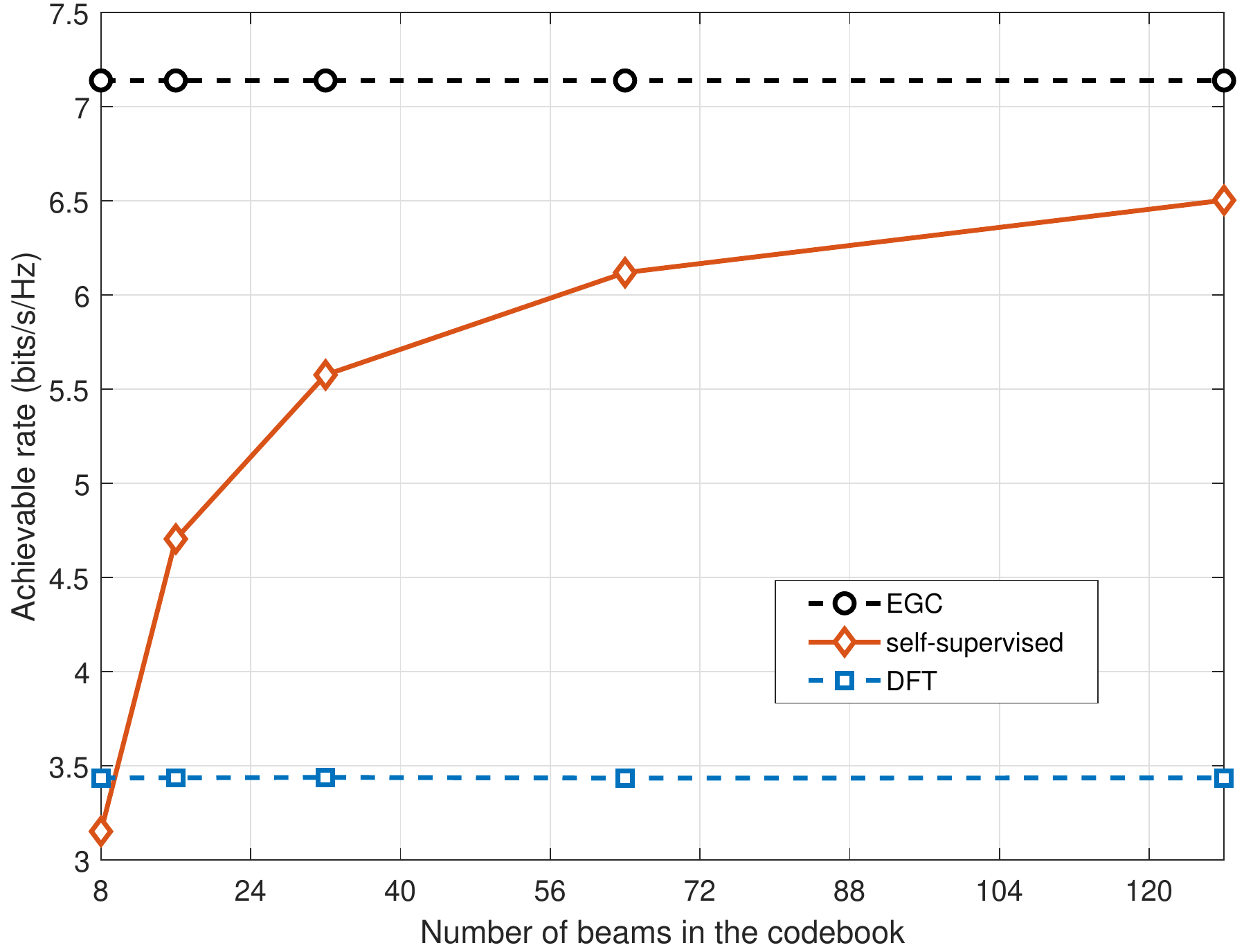}\label{online_impairment_ant0p1_ph0p5_noise_5dB_NLOS}}
	\caption{The achievable rate versus number of beams for the self-supervised solution. The performance is evaluated under 5 dB SNR, antenna spacing mismatch with $\sigma_d = 0.1 \lambda$ standard deviation, and phase mismatch with $\sigma_p = 0.4 \pi$. (a) shows the performance in LOS setting while (b) considers NLOS setting.}
\end{figure}

The performance of the self-supervised solution is evaluated under hardware impairments using the model introduced in Section \ref{sec:impairments}. The channels in datasets $\mathcal S^\rm{LOS}_{t_2},\mathcal S^\rm{NLOS}_{t_2}$, are corrupted with antenna spacing and phase mismatches that, respectively, have $\sigma_d = 0.1 \lambda$  and $\sigma_p = 0.4 \pi$  standard deviations. \figref{sup_online_impairment_noise_5dB_LOS} shows the simulation result of the proposed solution in a LOS setting. It maintains a similar performance to that presented in \figref{sup_online_noise_5dB_LOS} and displays an intriguing ability to combat the challenges imposed by the hardware impairments. \textbf{This indicates that the self-supervised solution can efficiently adapt to the corrupted (and arbitrary) array-response vectors, compensating for the antenna-spacing and phase mismatches}. The same performance can also be observed in the NLOS case. With the same hardware impairment settings, \figref{online_impairment_ant0p1_ph0p5_noise_5dB_NLOS} depicts the achievable rate versus the codebook size in a NLOS setting. The learned codebook continues to maintain the same trend as that in the LOS case. In fact, with 128 beams, the codebook learned can attain over $90\%$ of the achievable rate of the upper bound. Such ability is lacking in classical beam steering codebooks such as the DFT codebook. Compared to \figref{sup_online_noise_5dB_LOS} and \figref{sup_online_noise_5dB_NLOS}, the performance of DFT codebooks degrades significantly when impairments are present. The reason lies in the patterns of the corrupted array response vectors, which lose their directivity and experience critical distortion. 

\begin{figure}[t]
	\centering
	\subfigure[ ]{\includegraphics[width=.45\linewidth]{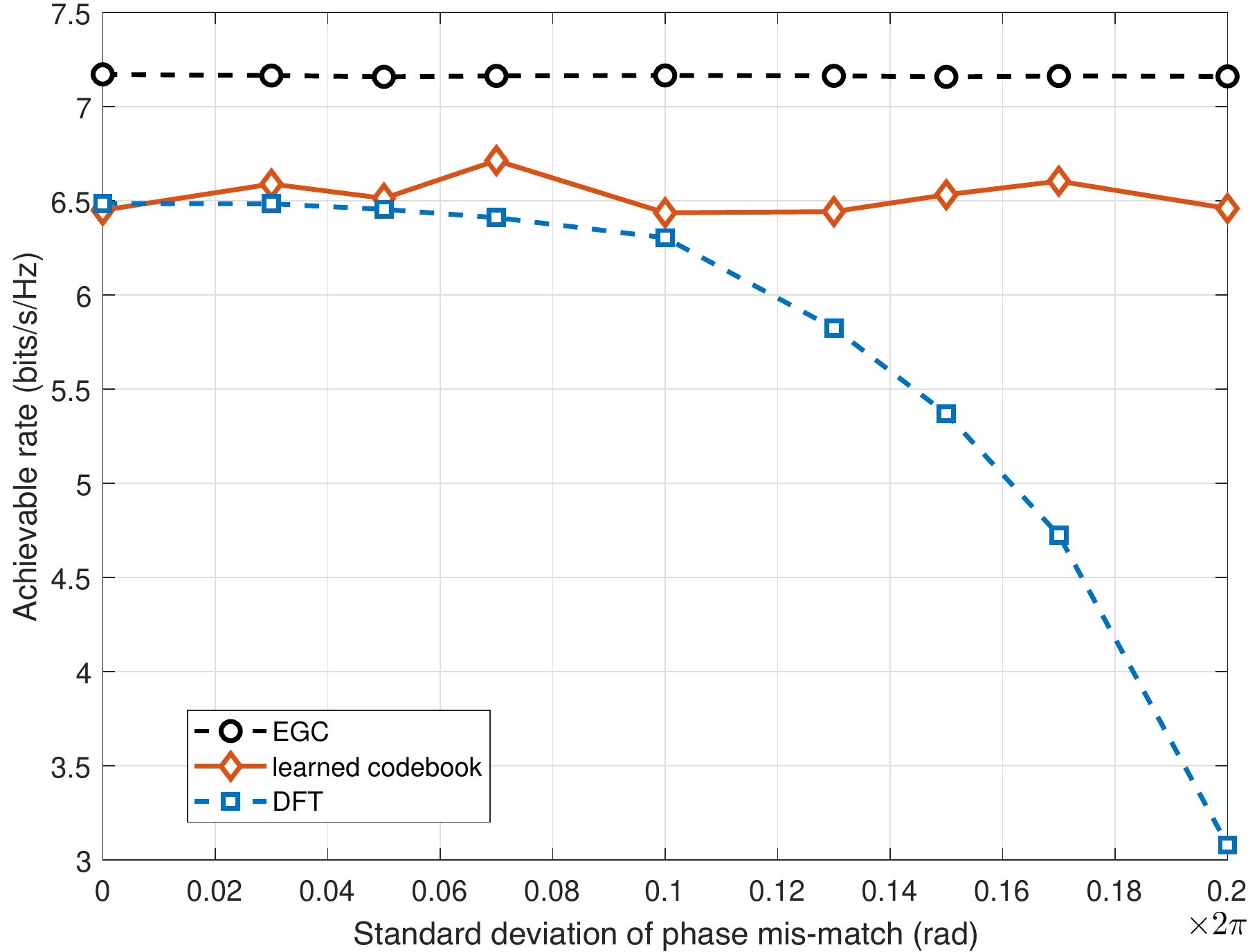}\label{online_ach_vs_std_of_phase_noise_5dB_LOS_NLOS_a}}
	\subfigure[ ]{\includegraphics[width=.45\linewidth]{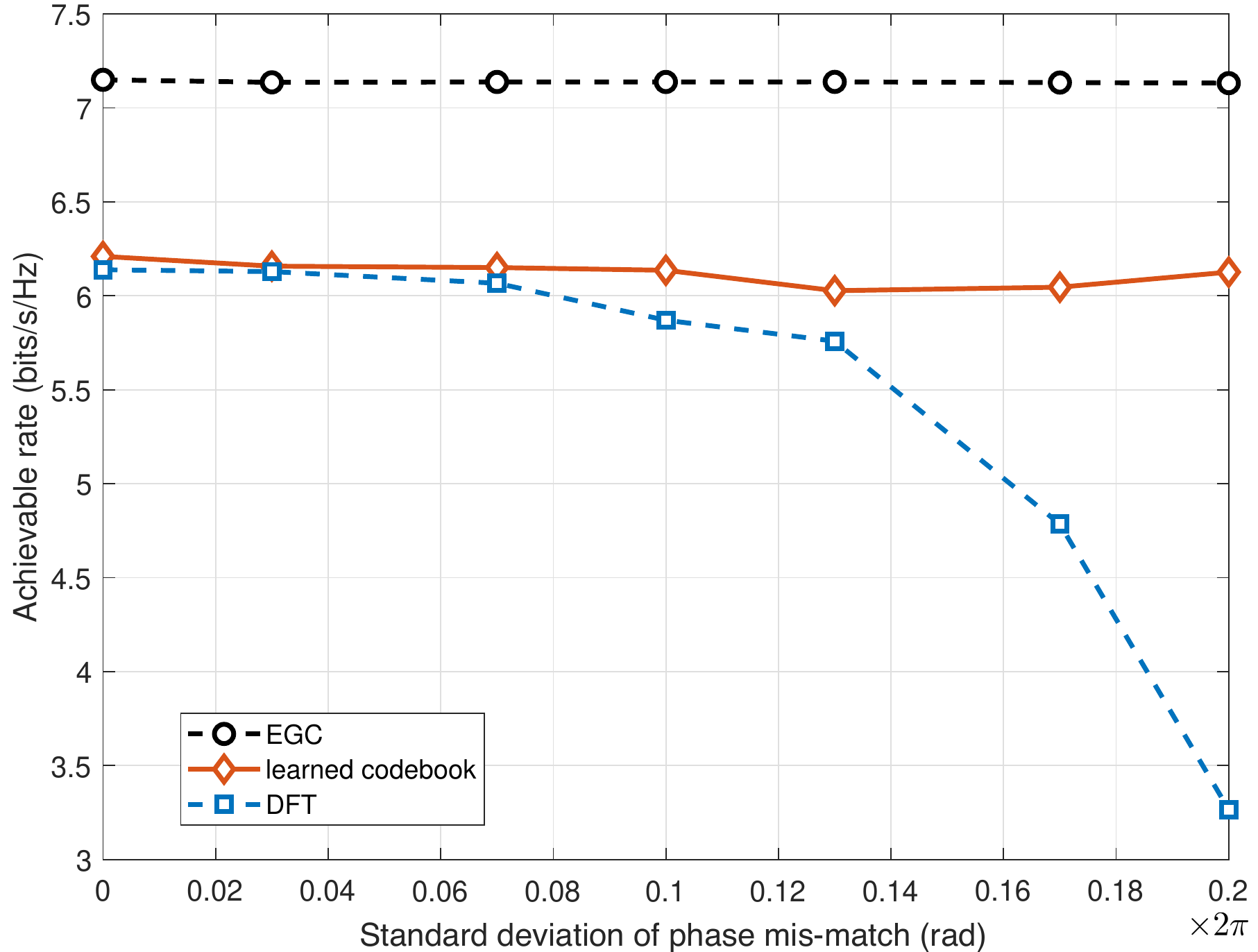}\label{online_ach_vs_std_of_phase_noise_5dB_LOS_NLOS_b}}
	
	\caption{The achievable rate vs. the standard deviation of phase mismatch with self-supervised solution in: (a) LOS and (b) NLOS settings. The performance is evaluated under 5dB receive SNR, antenna spacing mismatch with standard deviation of $0.1 \lambda$, and a 64-beams codebook.}
	\label{online_ach_vs_std_of_phase_noise_5dB_LOS_NLOS}
\end{figure}

It is important at this stage to pose the following question: How robust is the self-supervised solution? The answer to that question would provide some perspective on the capacity of the proposed solution to endure hardware impairments. In \figref{online_ach_vs_std_of_phase_noise_5dB_LOS_NLOS_a}, we plot the achievable rates  versus the standard deviation of the phase mismatch. The figure considers a LOS setting and a fixed antenna spacing mismatch with a standard deviation of $0.1 \lambda$. As the standard deviation of the phase increases, the self-supervised solution keeps a balanced performance. The DFT codebook, on the other hand, degrades drastically as the level of corruption increases. \textbf{This behavior demonstrates the  robustness of the proposed codebook learning approach and its ability to adapt to the various hardware impairments.} The same test with the same antenna spacing mismatch is repeated but in the NLOS setting, and the performance is shown in \figref{online_ach_vs_std_of_phase_noise_5dB_LOS_NLOS_b}. The proposed solution exhibits a similar performance to that in the LOS setting, which further emphasizes the conclusion on its robustness.

\begin{figure}[t]
	\centering
	\subfigure[ ]{\includegraphics[width=.3\linewidth]{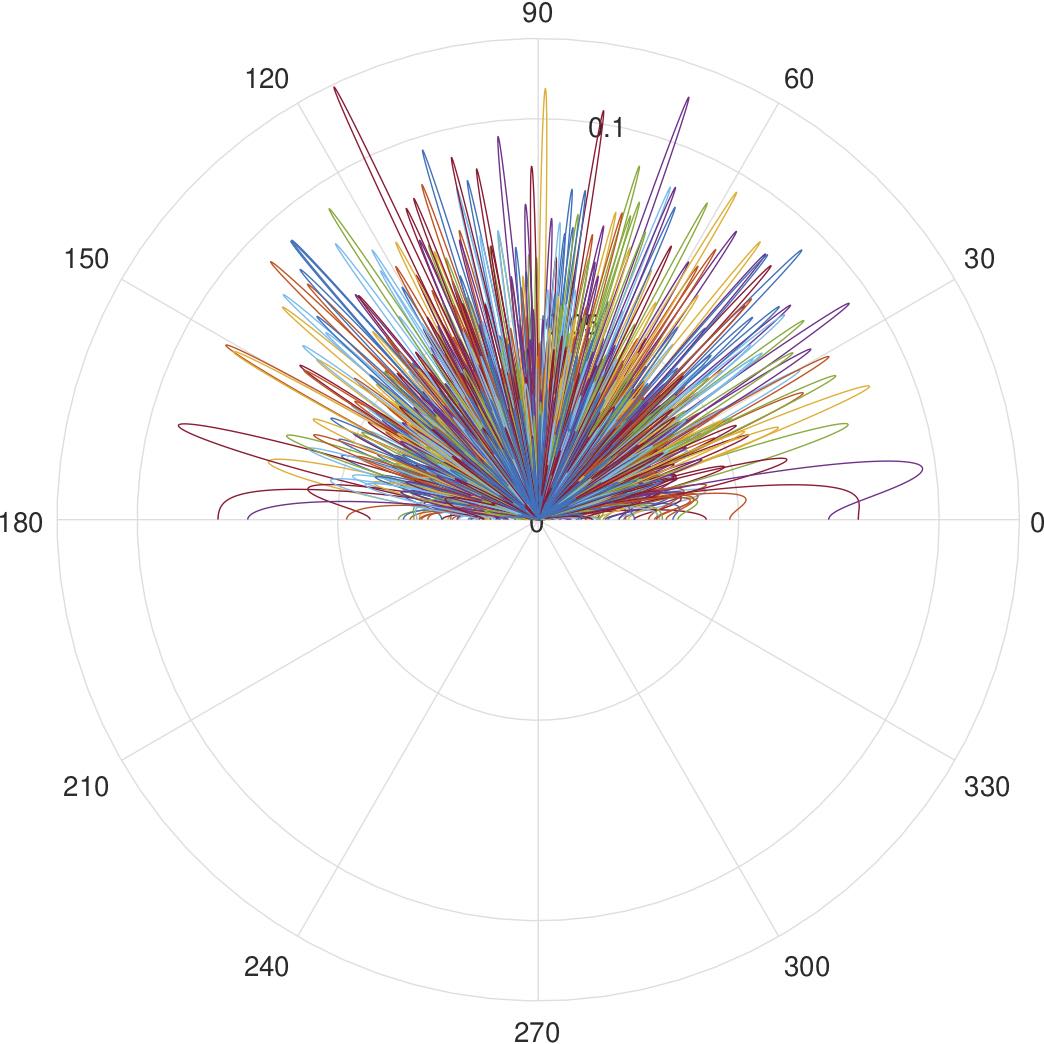}\label{all_patterns}}
	\subfigure[ ]{\includegraphics[width=.3\linewidth]{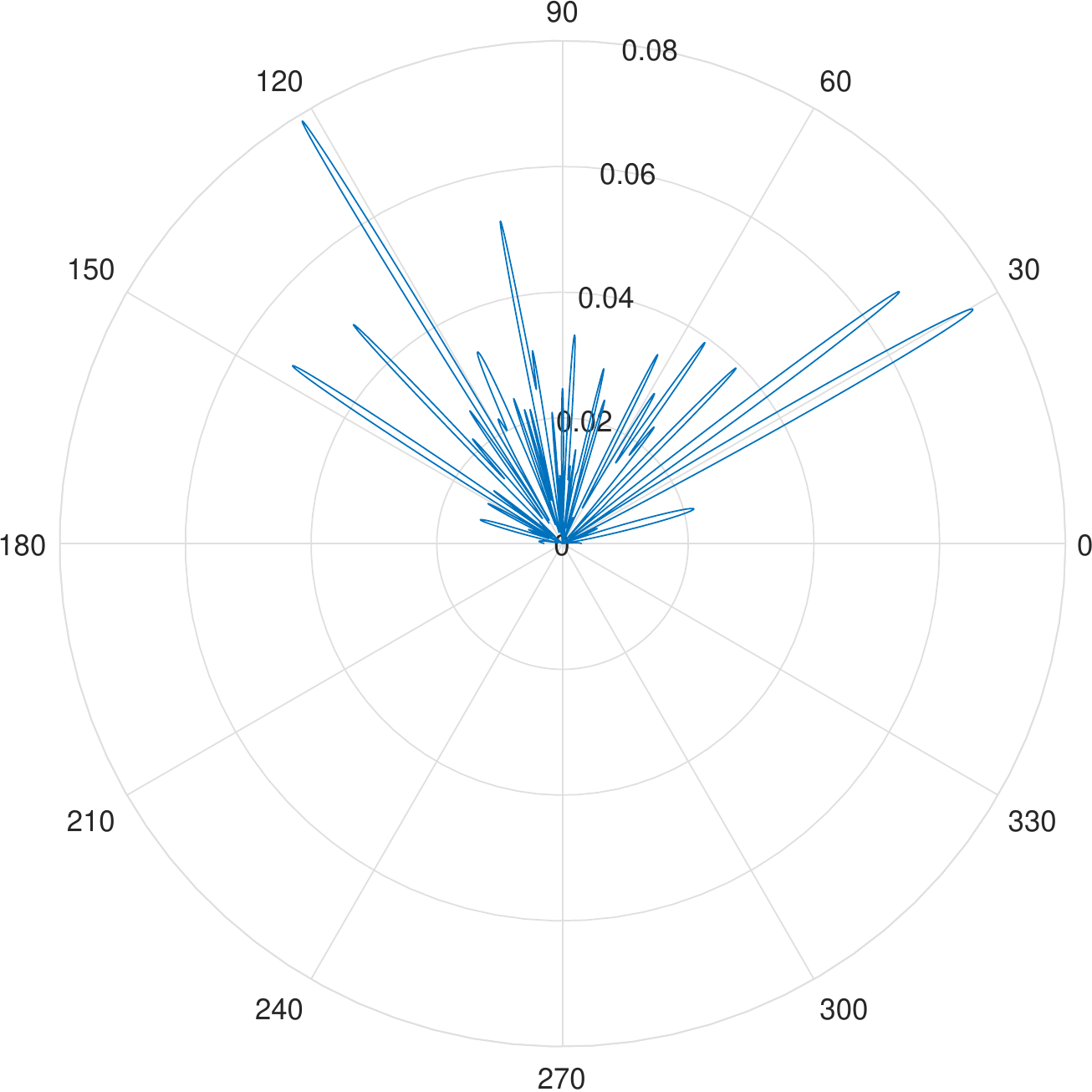}\label{single_beam}}
    \subfigure[ ]{\includegraphics[width=.3\linewidth]{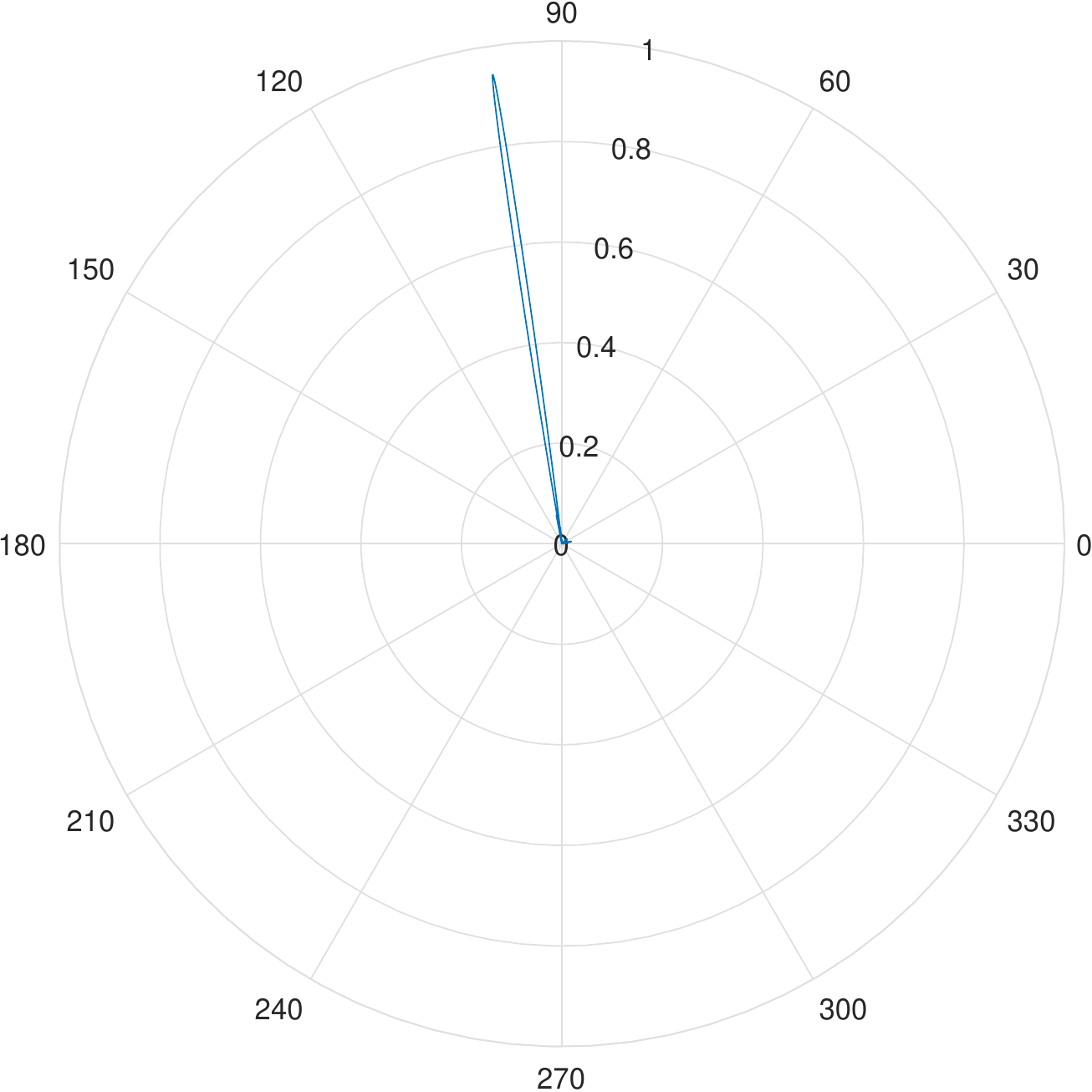}\label{projected_beam}}
	\caption{Beam patterns for codebook with 64 beams learned by the self-supervised solution in LOS setting with hardware impairments (antenna spacing and phase mismatches with, respectively, $0.1 \lambda$ and $0.4 \pi$ standard deviations). (a) shows all 64 beams (if plotted for a uniform array), (b) shows one of the codebook beams (if plotted for a uniform array), and, finally, (c) shows the same beam in (b) when plotted for the corrupted array (i.e., the actual beam pattern out of the corrupted array).}
	\label{self_sup_64_and_single_beam}
\end{figure}

To visualize what the proposed solution is learning exactly, \figref{self_sup_64_and_single_beam} plots different beam patterns from a learned codebook with hardware impairments and in a LOS setting. The first figure on the left, namely \figref{all_patterns}, shows all beam patterns in the learned codebook when projected on the angular space of the \textit{uniform} (uncorrupted) arrays. One of those beams in plotted again separately in \figref{single_beam}. While these beams appear distorted with multiple lobes, they actually look this way because they match the hardware impairments and mismatches. To prove that, we plotted the selected beam again in \figref{projected_beam} when projecting it on the angular space of the corrupted beams. In other words, this is the actual far-field beam pattern that the corrupted array will generate. This beam is clearly depicting the supposed pattern,  which is a single-lobe beam pointing to the user's direction. All that verifies the interesting capability of the proposed solution in learning beams that adapt to the surrounding environment and given hardware. 
%

\section{Conclusion}\label{concl}

In this paper, we considered hardware-constrained mmWave massive MIMO systems and developed a machine learning framework that learns environment and hardware aware beam codebooks. Achieving that was through designing novel complex-valued neural network architectures that use the neuron weights to  directly model the beamforming weights of the phase shifters. Further, these architectures account for the key hardware constraints such as  the constant-modulus and quantized-angles constraints. The proposed model is trained online in a self-supervised manner, avoiding the need for explicit channel state information. The developed approach was extensively evaluated using the publicly available dataset, DeepMIMO, at both LOS and NLOS environments. Simulation results show that the developed solution can learn codebook beams that adapt to the surrounding environment and user distribution, which can significantly  reduce the training overhead and improve the achievable data rates. Further, the results demonstrated the capability of the proposed solution in adapting the beam patterns to the given hardware impairments and array geometry. This highlights the potential gains of leveraging machine learning to develop deployment and hardware aware beamforming codebooks.


\appendix
\section{}
\subsection{Complex Differentiability}
\label{app:comp_diff}
The problem with \eqref{chain_rule} and \eqref{chain_rule_2} lie in their complex differentiability, more specifically, the complex differentiability of $\frac{\partial{\bf q}}{\partial{\bf z}}$ and $\frac{\partial{\bf z}}{\partial{\bf \theta_n}}$. We refer to the work of \cite{Trabelsi2017} where an argument is presented to circumvent this limitation. It states that in order to perform backpropagation in a complex-valued neural network, a sufficient condition is to have a cost function and activations that are differentiable with respect to the real and imaginary parts of each complex parameter in the network. Formally, let $w = w^\rm{r} + jw^\rm{im}\in \mathbb C$ and $z = f(w) \in \mathbb R$ such that it does not satisfy Cauchy-Riemann equations. In this case, $z$ is not complex differentiable, and the suggested way around this problem is to view $w^\rm{r}$ and $w^\rm{im}$ as two independent variables such that $w^\rm{r},\ w^\rm{im}\in \mathbb R$. Then, the ``gradient'' of $z$ is defined as
\begin{equation}
  \nabla z = \left[ \frac{\partial}{\partial w^\rm{r}}f(w), \frac{\partial}{\partial w^\rm{im}}f(w) \right]^T.
\end{equation}
For instance, if $z = (w^\rm{r})^2 + (w^\rm{im})^2 \in \mathbb R$, then
\begin{align}
  \nabla z & = \left[ \frac{\partial}{\partial w^\rm{r}}\left[(w^\rm{r})^2 + (w^\rm{im})^2)\right], \frac{\partial}{\partial w^\rm{im}}\left[(w^\rm{r})^2 + (w^\rm{im})^2)\right] \right]^T = 2\left[ w^\rm{r}, w^\rm{im} \right]^T
\end{align}

\subsection{Computing the Partials}\label{app:comp_der}
Going back to \eqref{chain_rule} and \eqref{chain_rule_2}, the factors $\frac{\partial \mathbf q}{\partial \mathbf z}$ and $\frac{\partial \mathbf z}{\partial \mathbf \theta_n}$ satisfy the condition, and, hence, we construct the Jacobian $\frac{\partial{\bf q}}{\partial{\bf z}}$ as
\begin{equation}\label{parQparZ}
  \frac{\partial{\bf q}}{\partial{\bf z}}
  = \left[\begin{array}{cccccccccccc}
            \frac{\partial q_1}{\partial z_{1}^\rm{r}} & 0 & 0 & \dots & 0 & 0 & \frac{\partial q_1}{\partial z_{1}^\rm{im}} & 0 & 0 & \dots & 0 & 0 \\
            0 & \frac{\partial q_2}{\partial z_{2}^\rm{r}} & 0 & \dots & 0 & 0 & 0 & \frac{\partial q_2}{\partial z_{2}^\rm{im}} & 0 & \dots & 0 & 0 \\
            \vdots & \vdots & \vdots & \ddots & \vdots & \vdots & \vdots & \vdots & \vdots & \ddots & \vdots & \vdots \\
            0 & 0 & 0 & \dots & 0 & \frac{\partial q_N}{\partial z_{N}^\rm{r}} & 0 & 0 & 0 & \dots & 0 & \frac{\partial q_N}{\partial z_{N}^\rm{im}}
          \end{array}\right]_{N\times2N}.
\end{equation}
The sparsity of the Jacobian follows from the fact that $\mathbf q$ is the result of an element-wise operation, see \eqref{powOut}. The reason behind its shape, i.e., $N\times 2N$ matrix, will be explained shortly. Since the output of the $n$th combiner $z_n$ is only determined by the $n$th column of the matrix $\mathbf W$ (see \eqref{fc_1}) and since the $n$th column of $\bf W$ is only a function in $\boldsymbol{\theta_n}$ (see \eqref{eq:ph_to_weight}), we can write the other Jacobian, namely 
$\frac{\partial{\bf z}}{\partial\boldsymbol{\theta}_n}$, as

\begin{equation}\label{parZparT}
  \frac{\partial{\bf z}}{\partial\boldsymbol{\theta}_n}
  = \left[\begin{array}{ccccccccc}
            0 & \dots & \frac{\partial z_{n}^\rm{r}}{\partial \theta_{n1}} & \dots & 0 & \dots & \frac{\partial z_{n}^\rm{im}}{\partial \theta_{n1}} & \dots & 0 \\
            0 & \dots & \frac{\partial z_{n}^\rm{r}}{\partial \theta_{n2}} & \dots & 0 & \dots & \frac{\partial z_{n}^\rm{im}}{\partial \theta_{n2}} & \dots & 0 \\
            \vdots & \ddots & \vdots & \ddots & \vdots & \ddots & \vdots & \ddots & \vdots \\
            0 & \dots & \frac{\partial z_{n}^\rm{r}}{\partial \theta_{nM}} & \dots & 0 & \dots & \frac{\partial z_{n}^\rm{im}}{\partial \theta_{nM}} & \dots & 0 \\
          \end{array}\right]^T_{M\times2N}.
\end{equation}
Now, to calculate $\frac{\partial z_{n}^\rm{r}}{\partial \theta_{n m}}$ or $\frac{\partial z_{n}^\rm{im}}{\partial \theta_{n m}}$, $\forall m = \left\{1, \dots, M\right\}$, we recall \eqref{fc_2} and write $z_{n}^\rm{r}$ and $z_{n}^\rm{im}$ as functions of the $n$th column of $\bf W$ as follows
\begin{align}
  z_n^\rm{r} =& \sum_{m=1}^{M} w_{n m}^\rm{r} h_m^\rm{r} - w_{n m}^\rm{im} h_m^\rm{im}, \label{zReal} \\
  z_n^\rm{im} =& \sum_{m=1}^{M} \left(-w_{n m}^\rm{im}\right) h_m^\rm{r} + w_{n m}^\rm{r} h_m^\rm{im}, \label{zImag}
\end{align}
where
\begin{equation}\label{euler}
  w_{n m}^\rm{r} = \cos\left(\theta_{n m}\right), ~ w_{n m}^\rm{im} = \sin\left(\theta_{n m}\right).
\end{equation}
The partials now are computed as follows
\begin{align}
  \frac{\partial z_{n}^\rm{r}}{\partial \theta_{n m}} =& \frac{\partial z_{n}^\rm{r}}{\partial w_{n m}^\rm{r}}\cdot \frac{\partial w_{n m}^\rm{r}}{\partial \theta_{n m}}
  + \frac{\partial z_{n }^\rm{r}}{\partial w_{n m}^\rm{im}}\cdot \frac{\partial w_{n m}^\rm{im}}{\partial \theta_{n m}}, \label{parzpartR} \\
  =& -h_m^\rm{r}\sin\left(\theta_{n m}\right) + h_m^\rm{im}\cos\left(\theta_{n m}\right), \label{expandZr}
 \end{align}
 and
 \begin{align}
  \frac{\partial z_{n}^\rm{im}}{\partial \theta_{n m}} =& \frac{\partial z_{n}^\rm{im}}{\partial \left(-w_{n m}^\rm{im}\right)}\cdot \frac{\partial \left(-w_{n m}^\rm{im}\right)}{\partial \theta_{n m}}
  + \frac{\partial z_{n}^\rm{im}}{\partial w_{n m}^\rm{r}}\cdot \frac{\partial w_{n m}^\rm{r}}{\partial \theta_{n m}}, \label{parzpartI} \\
  =& h_m^\rm{r}\cos\left(\theta_{n m}\right) + h_m^\rm{im}\sin\left(\theta_{n m}\right). \label{expandZi}
\end{align}
Evaluating \eqref{expandZr} and \eqref{expandZi} clearly relies on the channel estimates. This should not be a problem for the supervised solution, but for the self-supervised solution, the estimate obtained using \eqref{estCh}, namely $\widehat{\mathbf h}$, is substituted for $\mathbf h$.

Having found the partials, the reason behind the choice of the matrix shapes in \eqref{parQparZ} and \eqref{parZparT} could be explained. The final objective of \eqref{chain_rule} and \eqref{chain_rule_2} is to propagate back the error signal and update the parameters of the codebook as in \eqref{eq:update_rule}. The matrix forms of \eqref{parQparZ} and \eqref{parZparT} guarantees that the computation of $\frac{\partial\mathcal L}{\partial \boldsymbol{\theta}_n}$ could be performed in simple matrix multiplication, which is critical for efficient implementation.

\linespread{1.2}

\end{document}